# On Structuring Proof Search
# for First Order Linear Logic

Paola Bruscoli and Alessio Guglielmi

Technische Universität Dresden

Hans-Grundig-Str. 25 - 01062 Dresden - Germany

Paola.Bruscoli@Inf.TU-Dresden.DE, Alessio.Guglielmi@Inf.TU-Dresden.DE

**Abstract**   *Full first order linear logic can be presented as an abstract logic programming language in Miller's system* Forum*, which yields a sensible operational interpretation in the 'proof search as computation' paradigm. However,* Forum *still has to deal with syntactic details that would normally be ignored by a reasonable operational semantics. In this respect,* Forum *improves on Gentzen systems for linear logic by restricting the language and the form of inference rules. We further improve on* Forum *by restricting the class of formulae allowed, in a system we call* G-Forum*, which is still equivalent to full first order linear logic. The only formulae allowed in* G-Forum *have the same shape as* Forum *sequents: the restriction does not diminish expressiveness and makes* G-Forum *amenable to proof theoretic analysis.* G-Forum *consists of two (big) inference rules, for which we show a cut elimination procedure. This does not need to appeal to finer detail in formulae and sequents than is provided by* G-Forum*, thus successfully testing the internal symmetries of our system.*

## 1    Introduction

Forum [8, 9] is a presentation of linear logic, conceived by Miller and based on previous work by Andreoli [1], which only produces uniform proofs [10]. This guarantees that a sensible computational interpretation of proof search is possible. Surprisingly, Forum is complete for linear logic; this contrasts with the situation in classical logic, where a complete presentation that only produces uniform proofs is not possible [11]. Given linear logic's flexibility in interpreting a broad range of computational situations, Forum represents a major step towards practical applications.

This paper is motivated by the search for adequate operational models of Forum, especially behavioural models like labelled event structures, which describe causal relations between events. These are particularly important for some of the domains of application of Forum, namely the modelling of computations in concurrency and planning.

A proof in Forum mainly consists of small, mostly deterministic steps, corresponding to applying each of several inference rules. This determinism is not a surprise, of course, since Forum has been designed precisely for the purpose of reducing and isolating non-determinism. However, most of these steps do not correspond to 'interesting' observations in a computation. For example, two applications of a rule for $\bindnasrepma$ are necessary for decomposing $A \bindnasrepma (B \bindnasrepma C)$ into its constituents $A$, $B$, $C$; moreover, the order of application of rules would be different for decomposing $(A \bindnasrepma B) \bindnasrepma C$, but the result would be the same. Since in such a case one is interested only in the result $A, B, C$, the detail about the order of applications of the rule for $\bindnasrepma$ is not necessary. This is a trivial case of a more general phenomenon in which a result is essentially deterministic, but its computation depends on irrelevant factors like a casual decision about associations of formulae connected by $\bindnasrepma$ connectives. In other words, one might be interested in identifying derivations that differ for such details as the ones shown above.

To get Forum, Miller imposed certain restrictions on the sequents, inference rules, and possible connectives of linear logic, but he left formula building free. In this paper, we restrict the class of formulae allowed, along lines already imagined by Miller in [8], and we design correspondingly a system called G-Forum. By doing this, we have that formulae drive the construction of proofs in a very structured way, which allows us to individuate big chunks of derivations that essentially behave in a deterministic way: these will be the building blocks of our desired behavioural semantics. The restriction on formulae makes them isomorphic to the sequents in Forum. Thus, we obtain two results: we get derivations which are closer to the operational properties we want to observe, and we also get a clean correspondence between the object level (the language of formulae) and the meta level (sequents). At this point one wonders whether G-Forum has any good proof-theoretic qualities, and this is the main subject of this paper. We test the internal harmony of G-Forum the classical way: we show a cut elimination procedure for our system, and we show that it is not necessary to resort to Forum for it to work: the bigger granularity of G-Forum rules is enough.

We leave the development of the operational semantics to a future paper (the reader can consult [6] for some preliminary results); in this paper we limit ourselves to defining G-Forum and show cut elimination for it. In Sect. 2 we give a quick account of Forum, then we develop G-Forum in Sect. 3. The cut elimination proof is in Sect. 4.

## 2    First Order Forum

This section is a quick account of Miller's [8, 9], restricted to the first order case and with some minor technical and notational changes.

We deal with first order formal systems, and the following conventions apply.

**2.1    Notation**   The letters $h$, $k$ and $l$ denote *natural numbers*, and $i$ and $j$ are used as *indices* on natural numbers. *Multisets* are denoted by braces as in $\{\dots\}_+$; *multiset union* is $\uplus$ and the *empty multiset* is $\varnothing_+$. If $a$ belongs to multiset $M$ we write $a \in M$.

**2.2    Definition**   *First order variables* are denoted by $x$, $y$ and $z$; *terms* are denoted by $t$, *atoms* by $a$, $b$, $c$, …, $a(t_1, \dots, t_h)$, $b(\dots)$, $c(\dots)$, …. *Sequences* are denoted in vector notation, as in $\forall \vec{x}.a(\vec{t})$. *Formulae* are denoted by $F$ and other letters which will be introduced later on. Formulae are considered equal under $\alpha$-conversion.

This work is founded on linear logic; we are mainly interested in its first order sequent calculus presentation. We refer to the literature for details, especially to Girard's [5].

**2.3    Definition**   The formal system of full *first order linear logic*, in its Gentzen's sequents presentation, and its language, are both denoted by FOLL. Formulae in FOLL are freely built from first order atoms and *constants* $1$, $\bot$, $\top$, $0$ by using *binary connectives* $\otimes$, $\invamp$, $\&$, $\oplus$, $\multimap$, *modalities* $!$, $?$, *negation* $^\perp$ and the *quantifiers* $\forall$ and $\exists$. Constants $1$, $\bot$ and connectives $\otimes$, $\invamp$ and $\multimap$ are called the *multiplicatives*; $\top$, $0$, $\&$ and $\oplus$ are called the *additives*. *Equivalence* is written $\equiv$. In linear logic $F \equiv F'$ iff $(F \multimap F') \& (F' \multimap F)$ is provable.

Intuitionistic implication $\Rightarrow$ admits the well-known decomposition $F \Rightarrow F' \equiv {!F} \multimap F'$; we can consider $\Rightarrow$ part of our language.

**2.4    Definition**   The *binary connective* $\Rightarrow$ is introduced such that $F \Rightarrow F'$ is equivalent to $!F \multimap F'$.

**2.5    Definition**   Multiplicative connectives, except for $\multimap$, take precedence over additive ones; implications are the weakest connectives; modalities and quantifiers are stronger than binary connectives; negation takes precedence over everything. Implications associate to the right. Whenever possible, we omit parentheses.

For example, $!\forall x.a^\perp \multimap b \& c \invamp d \Rightarrow e$ stands for $\big(!(\forall x.(a^\perp))\big) \multimap \big((b \& (c \invamp d)) \Rightarrow e\big)$.

We briefly introduce the Forum formal system. The presentation corresponds to the one in [8], restricted to the first order case and with some minor modifications. An alternative and more detailed exposition can be found in [9].

**2.6   Definition**   The language of *first order* Forum is the subset of FOLL freely built over atoms and the constants $\perp$ and $\top$ by use of the binary connectives $\bindnasrepma$, $\&$, $\multimap$ and $\Rightarrow$ and of the quantifier $\forall$. We will say 'Forum' instead of 'first order Forum.' Generic Forum formulae are denoted by $A$ and $B$.

So, Forum presents fewer connectives than FOLL, by getting rid of some of the redundant ones. It is not difficult to prove the following equivalences in FOLL:

$$1 \equiv \perp^{\perp}, \qquad\qquad 0 \equiv \top^{\perp},$$
$$F \otimes F' \equiv (F^{\perp} \bindnasrepma F'^{\perp})^{\perp}, \qquad\qquad F \oplus F' \equiv (F^{\perp} \& F'^{\perp})^{\perp},$$
$$!F \equiv (F \Rightarrow \perp)^{\perp}, \qquad\qquad ?F \equiv F^{\perp} \Rightarrow \perp,$$
$$\exists x.F \equiv (\forall x.F^{\perp})^{\perp},$$
$$F^{\perp} \equiv F \multimap \perp.$$

Then, one can equivalently write any FOLL formula into the Forum language.

**2.7   Definition**   *Sequents* are expressions of the form

$$\begin{bmatrix} \Psi \\ \Gamma \end{bmatrix} A \vdash \begin{bmatrix} \\ \Lambda \end{bmatrix} \qquad \text{or} \qquad \begin{bmatrix} \Psi \\ \Gamma \end{bmatrix} \vdash \begin{bmatrix} \Xi \\ \Lambda \end{bmatrix},$$

where all formulae are Forum formulae and

- $\Psi$ is a finite multiset of formulae (the *left classical context* or *classical program*);
- $\Gamma$ is a finite multiset of formulae (the *left linear context* or *linear program*);
- $A$ is a formula (the *left focused formula*);
- $\Xi$ is a finite sequence of formulae (the *right linear context*);
- $\Lambda$ is a finite multiset of atoms (the *atomic context*).

$\Gamma$, $\Xi$ and $\Lambda$ are collectively referred to as the *linear context*. $\Psi$ and $\Gamma$ together are called the *program*. In the following $\Psi$, $\Gamma$, $\Xi$ and $\Lambda$ respectively stand for multisets, multisets and sequences of formulae and multisets of atoms. We write '$\Gamma, A$', or '$A, \Gamma$', instead of $\Gamma \uplus \{A\}_+$ and '$\Gamma, \Gamma'$' instead of $\Gamma \uplus \Gamma'$. Sequents are denoted by $\Sigma$. Sequents where no focused formula is present and $\Xi$ is empty are called *state sequents* and are written

$$\begin{bmatrix} \Psi \\ \Gamma \end{bmatrix} \vdash \begin{bmatrix} \\ \Lambda \end{bmatrix}.$$

**2.8   Definition**   An *inference rule* is an expression of the form $r \dfrac{\Sigma_1 \ \ldots \ \Sigma_h}{\Sigma}$, where $h \geqslant 0$, sequents $\Sigma_1, \ldots, \Sigma_h$ are the *premises* of the rule, $\Sigma$ is its *conclusion* and $r$ is the *name* of the rule. An inference rule with no premises is called an *axiom*.

**2.9   Definition**   Let Forum be the first order proof system defined by inference rule schemes in Fig. 1. Structural rules are: i (*identity*), $\mathsf{d_L}$ (*decide linear*), $\mathsf{d_C}$ (*decide classical*), a (*atom*). Logical rules, divided into left and right ones, are: $\perp_\mathsf{L}$, $\perp_\mathsf{R}$ (*bottom*); $\top_\mathsf{R}$ (*top*, there is no left rule); $\bindnasrepma_\mathsf{L}$, $\bindnasrepma_\mathsf{R}$ (*par*); $\&_\mathsf{LL}$, $\&_\mathsf{LR}$, $\&_\mathsf{R}$ (*with*); $\multimap_\mathsf{L}$, $\multimap_\mathsf{R}$ (*linear implication*); $\Rightarrow_\mathsf{L}$, $\Rightarrow_\mathsf{R}$ (*intuitionistic implication*); $\forall_\mathsf{L}$, $\forall_\mathsf{R}$ (*universal quantification*).

Consider proofs in Forum in a bottom-up reading. In the absence of a left focused formula, the right linear context is acted upon by right rules until it is empty; at that point a formula

**Structural Rules**

$$
\mathsf{i}\ \frac{}{\left[\begin{array}{c}\Psi\end{array}\right]a\vdash\left[\begin{array}{c}a\end{array}\right]}
\qquad
\mathsf{d_L}\ \frac{\left[\begin{array}{c}\Psi\\\Gamma\end{array}\right]A\vdash\left[\begin{array}{c}\Lambda\end{array}\right]}{\left[\begin{array}{c}\Psi,A\\\Gamma\end{array}\right]\vdash\left[\begin{array}{c}\Lambda\end{array}\right]}
\qquad
\mathsf{d_C}\ \frac{\left[\begin{array}{c}\Psi,A\\\Gamma\end{array}\right]A\vdash\left[\begin{array}{c}\Lambda\end{array}\right]}{\left[\begin{array}{c}\Psi,A\\\Gamma\end{array}\right]\vdash\left[\begin{array}{c}\Lambda\end{array}\right]}
\qquad
\mathsf{a}\ \frac{\left[\begin{array}{c}\Psi\\\Gamma\end{array}\right]\vdash\left[\begin{array}{c}\Xi\\a,\Lambda\end{array}\right]}{\left[\begin{array}{c}\Psi\\\Gamma\end{array}\right]\vdash\left[\begin{array}{c}a,\Xi\\\Lambda\end{array}\right]}
$$

**Left Rules**                                        **Right Rules**

$$
\bot_{\mathsf{L}}\ \frac{}{\left[\begin{array}{c}\Psi\end{array}\right]\bot\vdash\left[\begin{array}{c}\ \end{array}\right]}
\qquad\qquad
\bot_{\mathsf{R}}\ \frac{\left[\begin{array}{c}\Psi\\\Gamma\end{array}\right]\vdash\left[\begin{array}{c}\Xi\\\Lambda\end{array}\right]}{\left[\begin{array}{c}\Psi\\\Gamma\end{array}\right]\vdash\left[\begin{array}{c}\bot,\Xi\\\Lambda\end{array}\right]}
$$

$$
\top_{\mathsf{R}}\ \frac{}{\left[\begin{array}{c}\Psi\\\Gamma\end{array}\right]\vdash\left[\begin{array}{c}\top,\Xi\\\Lambda\end{array}\right]}
$$

$$
\Bumpeq_{\mathsf{L}}\ \frac{\left[\begin{array}{c}\Psi\\\Gamma\end{array}\right]A\vdash\left[\begin{array}{c}\Lambda\end{array}\right]\quad\left[\begin{array}{c}\Psi\\\Gamma'\end{array}\right]B\vdash\left[\begin{array}{c}\Lambda'\end{array}\right]}{\left[\begin{array}{c}\Psi\\\Gamma,\Gamma'\end{array}\right]A\,\Bumpeq\,B\vdash\left[\begin{array}{c}\Lambda,\Lambda'\end{array}\right]}
\qquad
\Bumpeq_{\mathsf{R}}\ \frac{\left[\begin{array}{c}\Psi\\\Gamma\end{array}\right]\vdash\left[\begin{array}{c}A,B,\Xi\\\Lambda\end{array}\right]}{\left[\begin{array}{c}\Psi\\\Gamma\end{array}\right]\vdash\left[\begin{array}{c}A\,\Bumpeq\,B,\Xi\\\Lambda\end{array}\right]}
$$

$$
\&_{\mathsf{LL}}\ \frac{\left[\begin{array}{c}\Psi\\\Gamma\end{array}\right]A\vdash\left[\begin{array}{c}\Lambda\end{array}\right]}{\left[\begin{array}{c}\Psi\\\Gamma\end{array}\right]A\,\&\,B\vdash\left[\begin{array}{c}\Lambda\end{array}\right]}
\quad
\&_{\mathsf{LR}}\ \frac{\left[\begin{array}{c}\Psi\\\Gamma\end{array}\right]B\vdash\left[\begin{array}{c}\Lambda\end{array}\right]}{\left[\begin{array}{c}\Psi\\\Gamma\end{array}\right]A\,\&\,B\vdash\left[\begin{array}{c}\Lambda\end{array}\right]}
\quad
\&_{\mathsf{R}}\ \frac{\left[\begin{array}{c}\Psi\\\Gamma\end{array}\right]\vdash\left[\begin{array}{c}A,\Xi\\\Lambda\end{array}\right]\quad\left[\begin{array}{c}\Psi\\\Gamma\end{array}\right]\vdash\left[\begin{array}{c}B,\Xi\\\Lambda\end{array}\right]}{\left[\begin{array}{c}\Psi\\\Gamma\end{array}\right]\vdash\left[\begin{array}{c}A\,\&\,B,\Xi\\\Lambda\end{array}\right]}
$$

$$
\multimap_{\mathsf{L}}\ \frac{\left[\begin{array}{c}\Psi\\\Gamma\end{array}\right]\vdash\left[\begin{array}{c}A\\\Lambda\end{array}\right]\quad\left[\begin{array}{c}\Psi\\\Gamma'\end{array}\right]B\vdash\left[\begin{array}{c}\Lambda'\end{array}\right]}{\left[\begin{array}{c}\Psi\\\Gamma,\Gamma'\end{array}\right]A\multimap B\vdash\left[\begin{array}{c}\Lambda,\Lambda'\end{array}\right]}
\qquad
\multimap_{\mathsf{R}}\ \frac{\left[\begin{array}{c}\Psi\\\Gamma,A\end{array}\right]\vdash\left[\begin{array}{c}B,\Xi\\\Lambda\end{array}\right]}{\left[\begin{array}{c}\Psi\\\Gamma\end{array}\right]\vdash\left[\begin{array}{c}A\multimap B,\Xi\\\Lambda\end{array}\right]}
$$

$$
\Rightarrow_{\mathsf{L}}\ \frac{\left[\begin{array}{c}\Psi\end{array}\right]\vdash\left[\begin{array}{c}A\end{array}\right]\quad\left[\begin{array}{c}\Psi\\\Gamma\end{array}\right]B\vdash\left[\begin{array}{c}\Lambda\end{array}\right]}{\left[\begin{array}{c}\Psi\\\Gamma\end{array}\right]A\Rightarrow B\vdash\left[\begin{array}{c}\Lambda\end{array}\right]}
\qquad
\Rightarrow_{\mathsf{R}}\ \frac{\left[\begin{array}{c}\Psi,A\\\Gamma\end{array}\right]\vdash\left[\begin{array}{c}B,\Xi\\\Lambda\end{array}\right]}{\left[\begin{array}{c}\Psi\\\Gamma\end{array}\right]\vdash\left[\begin{array}{c}A\Rightarrow B,\Xi\\\Lambda\end{array}\right]}
$$

$$
\forall_{\mathsf{L}}\ \frac{\left[\begin{array}{c}\Psi\\\Gamma\end{array}\right]A[t/x]\vdash\left[\begin{array}{c}\Lambda\end{array}\right]}{\left[\begin{array}{c}\Psi\\\Gamma\end{array}\right]\forall x.A\vdash\left[\begin{array}{c}\Lambda\end{array}\right]}
\qquad
\forall_{\mathsf{R}}\ \frac{\left[\begin{array}{c}\Psi\\\Gamma\end{array}\right]\vdash\left[\begin{array}{c}A[y/x],\Xi\\\Lambda\end{array}\right]}{\left[\begin{array}{c}\Psi\\\Gamma\end{array}\right]\vdash\left[\begin{array}{c}\forall x.A,\Xi\\\Lambda\end{array}\right]}
$$

where $y$ is not free in the conclusion

**Fig. 1**   *The first order* **Forum** *proof system*

becomes focused, in a $\mathsf{d_L}$ or $\mathsf{d_C}$ rule. Then left rules only are applicable, until new formulae reach the right linear context, through $\multimap_{\mathsf{L}}$ and $\Rightarrow_{\mathsf{L}}$ rules. Proofs in Forum are said to be *uniform* [10, 8, 9].

Our system's major differences with Forum as presented in [8] are: 1) our classical context is a multiset while in [8] it is a set; 2) our atomic context is a multiset while in [8] it is a sequence. These differences do not affect provability (and uniformity of proofs), as it can be proved trivially.

Representing derivations as directed trees whose nodes are sequents is typographically advantageous, especially in the cut elimination proof. The direction of the arrows corresponds to the tree growth during the search for a proof. It should be clear that there is no difference between our non-standard notation and the usual one.

**2.10  Definition**   To every instance of an inference rule $r \dfrac{\Sigma_1 \;\cdots\; \Sigma_h}{\Sigma}$, when $h > 0$, an *elementary derivation*

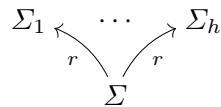

corresponds, i.e., a labelled directed tree whose root is labelled $\Sigma$, whose leaves are labelled $\Sigma_1$, ..., $\Sigma_h$ and whose arcs are labelled $r$; when $h = 0$ the corresponding elementary derivation is where $\circ$ is a mark distinct from every sequent. *Derivations* are non-empty, finite directed trees whose root is labelled by a sequent and whose other nodes are labelled by sequents or $\circ$ marks and such that every maximal subtree of depth 1 is an elementary derivation. Derivations are denoted by $\Delta$. Given a derivation $\Delta$, its *premises* are the labels of the leaves of $\Delta$ other than the $\circ$ ones; its *conclusion* is the sequent labelling the root of $\Delta$. A derivation $\Delta$ such that its premises are $\Sigma_1$, ..., $\Sigma_h$ and its conclusion is $\Sigma$ can be represented as

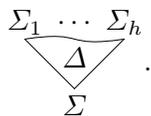

Sometimes the name of the derivation is not shown. If $\Delta$ is the derivation

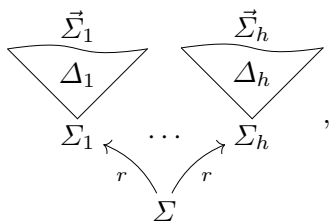

where $h \geq 0$, we define its *depth* $\mathsf{d}(\Delta)$ as the depth of the corresponding tree, i.e., $\mathsf{d}(\Delta) = \max\{\mathsf{d}(\Delta_1), \ldots, \mathsf{d}(\Delta_h)\} + 1$, where, for every sequent $\Sigma$, it holds $\mathsf{d}(\Sigma) = \mathsf{d}(\circ) = 0$. If $\Delta$ has no premises we say that $\Delta$ is a *proof*. Proofs are denoted by $\Pi$. We say that $\Pi$ *proves* (or *is a proof of*) its conclusion. We say that a formula $A$ is *provable* in Forum, or that Forum *proves* $A$, if a proof of $\left[\,\right] \vdash \left[A\right]$ exists.

For example, the premises of the derivation in Fig. 2 are $\left\{ \left[\,\right] a \vdash \left[a\right], \left[\,\right] a \vdash \left[a\right] \right\}_{+}$ and its conclusion is $\left[\,\right] a \,\⅋\, (b \,\⅋\, a) \vdash \left[a, a, b\right]$. This derivation can be completed into a proof by applying two identity rules to its premises.

Please note that arcs are not 'independent' in the growth process of a derivation: all arcs propagating from a node correspond to the application of the same inference rule.

By looking at Fig. 1 it is clear that if we make the classical context a set (as Miller does), derivability is not affected. In fact, the only impact is on the $\Rightarrow_{\mathsf{R}}$ rule, but things do not change, because the classical context is implicitly subject to weakening in all axioms.

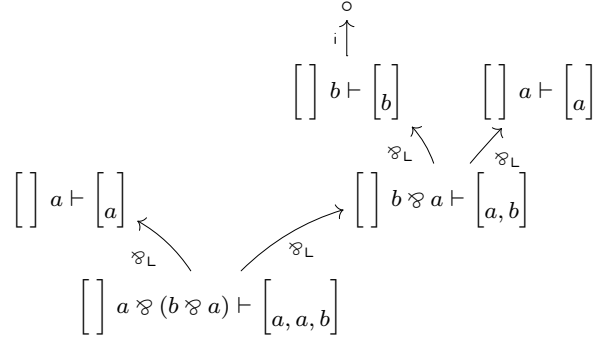

**Fig. 2**  *Example of derivation*

**2.11  Theorem**  *Every* Forum *formula is provable in* Forum *if and only if it is provable in* FOLL. (Miller [8, 9])

Since for every formula in FOLL an equivalent formula in Forum can be found, the Forum formal system can be used to prove formulae in FOLL.

# 3  Derivations at a Higher Level of Abstraction

Consider a formula $\delta = G_1 \Rightarrow \cdots \Rightarrow G_{k''} \Rightarrow H_1 \multimap \cdots \multimap H_{k'} \multimap a_1 \;\invamp\; \cdots \invamp a_k$. In Forum, in a bottom-up construction of a derivation, from $\left[\phantom{x}\right] \vdash \left[\delta\right]$ we are always led to the state sequents $\begin{bmatrix} G_1, \ldots, G_{k''} \\ H_1, \ldots, H_{k'} \end{bmatrix} \vdash \begin{bmatrix} a_1, \ldots, a_k \end{bmatrix}$. Let us call *clauses* formulae like $\delta$, where formulae $G_i$ and $H_j$, called *goals*, are of the form $\forall \vec{x}.(\delta_1 \;\&\; \cdots \;\&\; \delta_h)$, and where in the $\&$ conjunction only clauses are allowed.

In this section we derive a proof system equivalent to FOLL. The new proof system is in fact the old Forum proof system seen at a coarser abstraction level: rules are essentially macro derivations composed of many Forum rules, and the only formulae allowed are goals and clauses.

## 3.1  Goals and Clauses

We define goals and clauses, which are Forum formulae of a constrained shape; then we show that their language is equivalent to Forum and then to FOLL. We borrow from Miller the terminology on goals and clauses, and the reader should be aware that their use is more general than in standard logic programming, where clauses operate on goals in a clear hierarchical relation. In our formalism, goals and clauses are mutually recursive objects that only superficially bear a resemblance to goals and clauses of traditional logic programming.

### 3.1.1  Definition  Goals and clauses are recursively defined this way:

1)  A *goal* is a formula of the form

$$\forall \vec{x}.(\delta_1 \;\&\; \cdots \;\&\; \delta_h),$$

where $\vec{x}$ can be empty, $h \geqslant 0$ and every $\delta_i$ is a clause. When $h = 0$ a goal is $\forall \vec{x}.\top$.

2)  A *clause* $\delta$ is a formula of the form

$$G_1 \Rightarrow \cdots \Rightarrow G_{k''} \Rightarrow H_1 \multimap \cdots \multimap H_{k'} \multimap a_1 \;\invamp\; \cdots \invamp a_k,$$

where $k, k', k'' \geqslant 0$, formulae $G_i$ and $H_i$ are goals and formulae $a_i$ are atoms. Goals $G_i$ are called the *classical premises* of $\delta$, goals $H_i$ are its *linear premises* and $a_1 \,\bindnasrepma\, \cdots \,\bindnasrepma\, a_k$ is the *head* of the clause. We define $\mathsf{hd}(\delta) = \{a_1, \ldots, a_k\}_+$, $\mathsf{lp}(\delta) = \{H_1, \ldots, H_{k'}\}_+$ and $\mathsf{cp}(\delta) = \{G_1, \ldots, G_{k''}\}_+$. When $k = 0$ the head is $\bot$. When $k' = 0$ and $k'' = 0$ clauses assume the following special forms, respectively:

$$G_1 \Rightarrow \cdots \Rightarrow G_{k''} \Rightarrow a_1 \,\bindnasrepma\, \cdots \,\bindnasrepma\, a_k \quad \text{and}$$

$$H_1 \multimap \cdots \multimap H_{k'} \multimap a_1 \,\bindnasrepma\, \cdots \,\bindnasrepma\, a_k.$$

The letters $G$ and $H$ always denote goals and the letter $\delta$ always denotes clauses.

Clearly, a clause is also a goal.

The shape of goals and clauses is not due to chance, of course. On a technical level, it is motivated by the desire of keeping the cut elimination procedure *inside* the system we are going to define. This means that we would not consider acceptable eliminating cuts by resorting to the more primitive level of abstraction in which generic formulae and Forum inference rules are available. The only way to convince oneself of this is to try and modify the definition and see the impact on the cut elimination procedure. There are possibly many solutions to this problem, and the one we present here is probably only one of many.

However, there is a better explanation, which also offers a unique solution, the one we adopt: goals correspond to the shape of a proof tree, and clauses correspond to state sequents. In fact, a goal stands for the collection of branches of a tree, conveniently quantified universally (note that the branches of a derivation tree may share variables). Every branch of a derivation tree ends in a state sequent, as we will see, and this corresponds to a clause. The mutual recursion between goals and clauses correspond to the phases in the construction of a proof that we are going to explore in the rest of the paper. There is a certain mysticism in this correspondence, and we are not sure we really understand it enough; for the time being we content ourselves in seeing that it works. But let us now get back to the properties of goals and clauses.

In cut-free sequent systems enjoying the subformula property, like the one we are dealing with, various fragments, which differ in the connectives allowed, can be cut out of bigger ones, while maintaining provability unaffected in the corresponding languages. For example, we could take the fragment of Forum in which $\&$ is not allowed; since, reading proofs bottom-up, no rule can introduce connectives not already present in its conclusion, provability for formulae not containing $\&$ would not be influenced. There is sort of an independence, or modularity, among connectives, which we want to preserve, because it is a valuable property in language design.

We show two ways of getting equivalence between goals and clauses and generic formulae; the first one, with goals, respects independence of connectives by using in an essential way distributivity of $\,\bindnasrepma\,$ over $\&$.

**3.1.2 Theorem** *Every formula in* FOLL *is equivalent to a goal in* Forum.

**Proof** We already know that for every formula in linear logic there are equivalent formulae in Forum. We show that, taken any formula in Forum, we can exhibit an equivalent goal.

We use the following absorption equivalences:

1)  $F \,\bindnasrepma\, \bot \equiv F$.

2)  $F \,\bindnasrepma\, \top \equiv \top$.

3)  $F \,\&\, \top \equiv F$.

We also use the following equivalences:

4)  $F \,\bindnasrepma\, (F' \,\&\, F'') \equiv (F \,\bindnasrepma\, F') \,\&\, (F \,\bindnasrepma\, F'')$.

5)  $\forall x. F \,\bindnasrepma\, F' \equiv \forall x. (F \,\bindnasrepma\, F')$ whenever $x$ is not free in $F'$.

6)  $\forall x. F \,\&\, F' \equiv \forall x. (F \,\&\, F')$ whenever $x$ is not free in $F'$.

Let $A$ be a formula in Forum: the proof is by induction on its structure.

**Basis Cases**

- $A$ is an atom.
- $A = \bot$.
- $A = \top$.

In the cases above, $A$ is a goal.

**Inductive Cases**

Given $B$ and $B'$, by the induction hypothesis we suppose we are also given two goals $G$ and $G'$ such that

$$B \equiv G = \forall \vec{x}.(\delta_1 \,\&\, \cdots \,\&\, \delta_h),$$
$$B' \equiv G' = \forall \vec{y}.(\delta'_1 \,\&\, \cdots \,\&\, \delta'_{h'}),$$

where $\vec{x}$ and $\vec{y}$ may be empty and $h$ and $h'$ may be 0. The following cases may occur.

- $A = B \,\bindnasrepma\, B'$. By applications of equivalence 5 and renaming of bounded variables, if necessary, we get

$$A \equiv \forall \vec{z}.\big((\delta_1 \,\&\, \cdots \,\&\, \delta_h) \,\bindnasrepma\, (\delta'_1 \,\&\, \cdots \,\&\, \delta'_{h'})\big).$$

If $h = 0$ or $h' = 0$ we can conclude that $A \equiv \forall \vec{z}.\top$, by making use of equivalence 2. Otherwise, we may repeatedly apply equivalence 4 above, and we get:

$$A \equiv \forall \vec{z}.\Big(\big((\delta_1 \,\bindnasrepma\, \delta'_1) \,\&\, \cdots \,\&\, (\delta_h \,\bindnasrepma\, \delta'_1)\big) \,\&\, \cdots \,\&\, \big((\delta_1 \,\bindnasrepma\, \delta'_{h'}) \,\&\, \cdots \,\&\, (\delta_h \,\bindnasrepma\, \delta'_{h'})\big)\Big).$$

For $1 \leqslant i \leqslant h$ and $1 \leqslant j \leqslant h'$, let

$$\delta_i = G^i_1 \Rightarrow \cdots \Rightarrow G^i_{h''_i} \Rightarrow H^i_1 \multimap \cdots \multimap H^i_{h'_i} \multimap a^i_1 \,\bindnasrepma\, \cdots \,\bindnasrepma\, a^i_{h_i},$$
$$\delta'_j = G'^j_1 \Rightarrow \cdots \Rightarrow G'^j_{k''_j} \Rightarrow H'^j_1 \multimap \cdots \multimap H'^j_{k'_j} \multimap a'^j_1 \,\bindnasrepma\, \cdots \,\bindnasrepma\, a'^j_{k_j}.$$

Since $F \Rightarrow F' \equiv\, !F \multimap F'$ and $F \multimap F' \equiv F^\perp \,\bindnasrepma\, F'$, commutativity of $\bindnasrepma$ suffices to show that

$$\delta_i \,\bindnasrepma\, \delta'_j \equiv G^i_1 \Rightarrow \cdots \Rightarrow G^i_{h''_i} \Rightarrow G'^j_1 \Rightarrow \cdots \Rightarrow G'^j_{k''_j} \Rightarrow$$
$$H^i_1 \multimap \cdots \multimap H^i_{h'_i} \multimap H'^j_1 \multimap \cdots \multimap H'^j_{k'_j} \multimap$$
$$a^i_1 \,\bindnasrepma\, \cdots \,\bindnasrepma\, a^i_{h_i} \,\bindnasrepma\, a'^j_1 \,\bindnasrepma\, \cdots \,\bindnasrepma\, a'^j_{k_j}.$$

Special cases where $h_i = 0$ or $k_j = 0$ are handled by equivalence 1 above.

- $A = B \,\&\, B'$. By applications of equivalence 6 and renaming of bounded variables, if necessary, we get

$$A \equiv \forall \vec{z}.(\delta_1 \,\&\, \cdots \,\&\, \delta_h \,\&\, \delta'_1 \,\&\, \cdots \,\&\, \delta'_h).$$

If $h = 0$ or $h' = 0$ use equivalence 3.

- $A = B \multimap B'$. By using equivalences 5 and 4, and by renaming bounded variables if necessary, we have:

$$A \equiv G^\perp \,\bindnasrepma\, \forall \vec{y}.(\delta'_1 \,\&\, \cdots \,\&\, \delta'_h)$$
$$\equiv \forall \vec{z}.(G^\perp \,\bindnasrepma\, (\delta'_1 \,\&\, \cdots \,\&\, \delta'_h))$$
$$\equiv \forall \vec{z}.((G^\perp \,\bindnasrepma\, \delta'_1) \,\&\, \cdots \,\&\, (G^\perp \,\bindnasrepma\, \delta'_h)).$$

By commutativity of $\bindnasrepma$ it is easily seen that every $(G^\perp \,\bindnasrepma\, \delta'_i)$ is a clause. If $B' \equiv \top$ then $A \equiv \top$.

- $A = B \Rightarrow B'$. The argument goes as in the previous case.
- $A = \forall x.B$. Trivial.      $\square$

From the proof of the theorem we can derive an obvious algorithm that transforms a Forum formula into a goal. If $A$ is a Forum formula and $G$ the equivalent goal found by the algorithm, then the set of connectives appearing in $G$ is not greater than that of $A$. The translation could transform a linear logic formula or a generic Forum formula into a much bigger and perhaps obscure goal in Forum. We are going to see in a moment a more direct translation of Forum generic formulae into clauses that does not suffer from this problem. The translation does not respect independence of all connectives, because it introduces $\multimap$ and $\bot$, what should not be considered a severe constraint. First of all, we note:

**3.1.3 Corollary**    *Every formula in* FOLL *is equivalent to a clause.*

**Proof**    Let $F$ be a formula and $G \equiv F$, where $G$ is obtained as in Th. 3.1.2. Then, $(G \multimap \bot) \multimap \bot$ is a clause equivalent to $G$.    □

It is possible to prove the equivalence of Forum formulae and clauses in a shorter way, by using a double negation trick, as follows.

**3.1.4 Theorem**    *Every formula in* FOLL *is equivalent to a clause.*

**Proof**    Structural induction on a Forum formula $A$ equivalent to the given FOLL formula.

**Basis Cases**

If $A$ is an atom or $\bot$, just take $A$ as the equivalent clause. If $A = \top$ take $(\top \multimap \bot) \multimap \bot$.

**Inductive Cases**

Let $\delta$ and $\delta'$ be clauses equivalent to formulae $B$ and $B'$ in Forum, respectively.

- $A = B \parr B'$. Consider $\delta \parr \delta'$ and use the commutative property of $\parr$. Use $A \parr \bot \equiv A$ if necessary.
- $A = B \& B'$. Take $(\delta \& \delta' \multimap \bot) \multimap \bot$.
- $A = B \multimap B'$. Consider $\delta \multimap \delta' \equiv \delta^\bot \parr \delta'$ and use the commutative property of $\parr$.
- $A = B \Rightarrow B'$. Consider $\delta \Rightarrow \delta' \equiv (!\delta)^\bot \parr \delta'$ and use the commutative property of $\parr$.
- $A = \forall \vec{x}. B$. Take $(\forall \vec{x}. \delta \multimap \bot) \multimap \bot$.    □

One should be aware, though, that there are some concerns in Miller's [8] about clauses (similar to ours) with degenerate head $\bot$. Clauses of that kind, when at the left of $\vdash$, are always available to rewritings, what could be cause of explosion of the search space of proofs. How to translate formulae into goals and clauses is then a matter of careful judgment, to be exercised on the concrete situations one should deal with.

## 3.2    Deriving in the Right Context

We start here an analysis of the behaviour of goals and clauses in Forum. Our purpose is to isolate big chunks of derivations, whose shape is forced by the combined constraints of Forum inference rules and the restrictions on syntax we imposed in the previous subsection. We consider these big derivations as instances of (big) inference rules, whose operational meaning is reminiscent of traditional logic programming, but, of course, more general.

The analysis we perform is very straightforward: it only requires careful inspection of the rules. One way of looking at what we do here is the following: linear logic is a system with many rules, each of which performs a little operational task. Here, we head towards a system with only two rules, each of which has a somewhat complex behaviour. The point is that this behaviour is *manageable* in two senses: on one hand, it corresponds to a generalised view of logic programming, as intuitive as Miller's one in Forum (in our opinion); on the other hand, it is possible to define a cut elimination procedure for the two-rule formalism that, even when spelled out in full detail, as we do here, requires a comparable effort to getting cut elimination in unconstrained linear logic. In a subsequent paper we will see how these rules make sense in a third way, which is being in good

$$\begin{bmatrix} \Psi, G_1, \ldots, G_{k''} \\ \Gamma, H_1, \ldots, H_{k'} \end{bmatrix} \vdash \begin{bmatrix} & & & \Xi \\ a_1, \ldots, a_k, & \Lambda \end{bmatrix}$$

$$\text{\small(}\otimes_{\mathsf{R}} \text{ or } \mathtt{a})^\star \uparrow$$

$$\begin{bmatrix} \Psi, G_1, \ldots, G_{k''} \\ \Gamma, H_1, \ldots, H_{k'} \end{bmatrix} \vdash \begin{bmatrix} a_1 \otimes \cdots \otimes a_k, & \Xi \\ & \Lambda \end{bmatrix}$$

$$\multimap_{\mathsf{R}}^\star \uparrow$$

$$\begin{bmatrix} \Psi, G_1, \ldots, G_{k''} \\ \Gamma \end{bmatrix} \vdash \begin{bmatrix} H_1 \multimap \cdots \multimap H_{k'} \multimap a_1 \otimes \cdots \otimes a_k, & \Xi \\ & \Lambda \end{bmatrix}$$

$$\Rightarrow_{\mathsf{R}}^\star \uparrow$$

$$\begin{bmatrix} \Psi \\ \Gamma \end{bmatrix} \vdash \begin{bmatrix} G_1 \Rightarrow \cdots \Rightarrow G_{k''} \Rightarrow H_1 \multimap \cdots \multimap H_{k'} \multimap a_1 \otimes \cdots \otimes a_k, & \Xi \\ & \Lambda \end{bmatrix}$$

**Fig. 3**   *Clause reduction right inference rule* $\delta_{\mathsf{R}}$

correspondence to a sensible concurrent operational semantics. These should be reasons enough for the reader to struggle through the following, admittedly technical and tedious, definitions and propositions.

**3.2.1 Definition**   Let $\delta_{\mathsf{R}}$ be the following *clause reduction right* inference rule, shown in Fig. 3 in terms of Forum rules:

$$\delta_{\mathsf{R}} \frac{\begin{bmatrix} \Psi, \mathsf{cp}(\delta) \\ \Gamma, \mathsf{lp}(\delta) \end{bmatrix} \vdash \begin{bmatrix} & \Xi \\ \mathsf{hd}(\delta), & \Lambda \end{bmatrix}}{\begin{bmatrix} \Psi \\ \Gamma \end{bmatrix} \vdash \begin{bmatrix} \delta, \Xi \\ \Lambda \end{bmatrix}} \ .$$

In the figure $k > 0$ and $k', k'' \geqslant 0$. Starred inference rule names mean repeated application of the rule, or no application at all; $(\otimes_{\mathsf{R}} \text{ or } \mathtt{a})$ stands for 'application of one of either $\otimes_{\mathsf{R}}$ or $\mathtt{a}$.' In the special case where $k = 0$ the upper sequence of $(\otimes_{\mathsf{R}} \text{ or } \mathtt{a})$ rules is replaced by a single application of $\perp_{\mathsf{R}}$.

Clauses assume different meanings depending on whether they appear at the left or at the right of the entailment symbol $\vdash$. When a clause appears at the right of $\vdash$, we operationally interpret it as follows:

- classical premises are added to the classical program: they can be used at will (or not used at all) in the rest of the computation;
- linear premises are added to the linear program: they must be used exactly once in the rest of the computation;
- atoms in the head go into the atomic context: they are added to the current multiset of resources upon which the program will act.

We informally say that a $\delta_{\mathsf{R}}$ rule *loads* the contexts, by *reducing* clauses.

$\delta_{\mathsf{R}}$ is nothing more than a shortening for a piece of a derivation. The following proposition justifies its introduction.

**3.2.2 Proposition**   *Every proof of* $\begin{bmatrix} \Psi \\ \Gamma \end{bmatrix} \vdash \begin{bmatrix} \delta, \Xi \\ \Lambda \end{bmatrix}$ *has shape*

$$\begin{bmatrix} \Psi' \\ \Gamma' \end{bmatrix} \vdash \begin{bmatrix} \Xi \\ \Lambda' \end{bmatrix}$$

$$\delta_{\mathsf{R}} \uparrow$$

$$\begin{bmatrix} \Psi \\ \Gamma \end{bmatrix} \vdash \begin{bmatrix} \delta, \Xi \\ \Lambda \end{bmatrix}$$
.

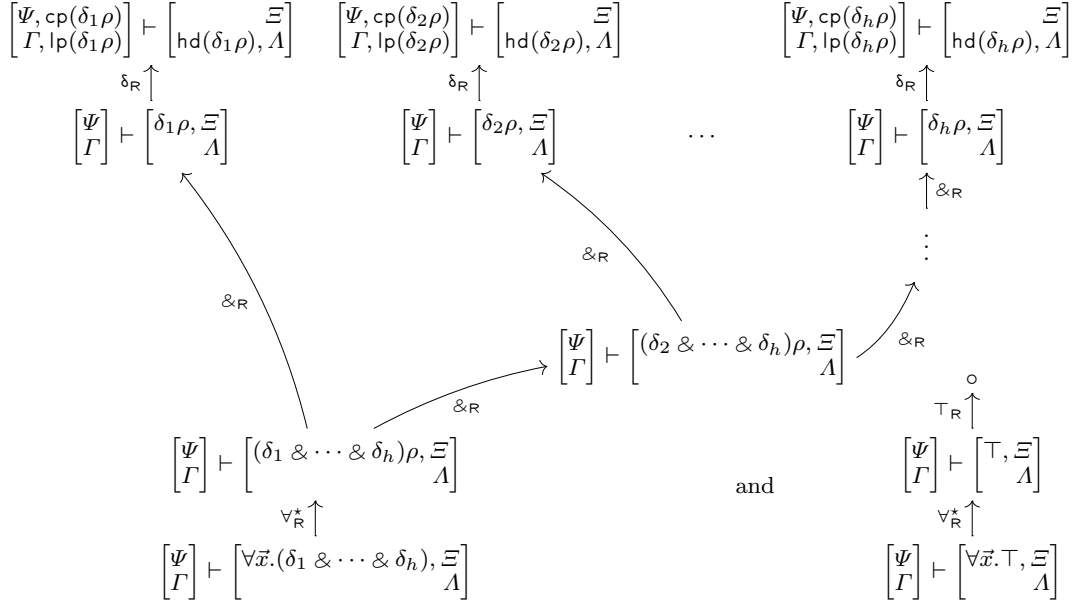

**Fig. 4** *Goal reduction right inference rule* $\mathsf{G_R}$ *when* $h > 0$ *and* $h = 0$

**Proof** By reasoning bottom-up, each application of an inference rule is compulsory. □

All Forum inference rules applied in $\delta_\mathsf{R}$ are right ones. This is of course an aspect of the fact that Forum produces only uniform proofs (see [8, 9, 10]). Rules $\top_\mathsf{R}$, $\&_\mathsf{R}$ and $\forall_\mathsf{R}$ are still missing: they will appear in the reduction of goals.

We can build on $\delta_\mathsf{R}$ an inference rule which reduces goals in the right linear context.

**3.2.3 Definition** Let $\mathsf{G_R}$ be the following *goal reduction right* inference rule, shown in Fig. 4 in terms of $\delta_\mathsf{R}$ and Forum rules:

$$\mathsf{G_R} \ \frac{\begin{bmatrix} \Psi, \mathsf{cp}(\delta_1\rho) \\ \Gamma, \mathsf{lp}(\delta_1\rho) \end{bmatrix} \vdash \begin{bmatrix} \Xi \\ \mathsf{hd}(\delta_1\rho), \Lambda \end{bmatrix} \quad \cdots \quad \begin{bmatrix} \Psi, \mathsf{cp}(\delta_h\rho) \\ \Gamma, \mathsf{lp}(\delta_h\rho) \end{bmatrix} \vdash \begin{bmatrix} \Xi \\ \mathsf{hd}(\delta_h\rho), \Lambda \end{bmatrix}}{\begin{bmatrix} \Psi \\ \Gamma \end{bmatrix} \vdash \begin{bmatrix} \forall \vec{x}.(\delta_1 \& \cdots \& \delta_h), \Xi \\ \Lambda \end{bmatrix}},$$

where $\vec{x}$ can be empty, $\rho$ is an appropriate renaming substitution and $h \geqslant 0$. In the figure only one choice among the possible associations of $\&$ connectives has been considered, but every choice leads to the same multiset of premises.

This whole reduction phase is deterministic: in the end a goal is reduced to pieces with no choice about the possible outcome, except for the rather immaterial choice of eigenvariables in $\mathsf{G_R}$ rules. The Forum system has been designed to reduce choices to a minimum, in a bottom-up construction of a proof. Still, some 'not necessary' sequentialisation exists: in the case above it resides in the binary treatment of associative connectives. We can consider the $\mathsf{G_R}$ rule at the abstraction level in which all premises are reached at the same time in a parallel way, thus hiding the sequentialisation at the Forum's level of abstraction. In other words we can consider every instance of the $\mathsf{G_R}$ rule a representative of an equivalence class of derivations, differing only in the associations of $\&$ connectives.

We can perform on the $\mathsf{G_R}$ rule the same kind of simple reasoning we did for $\delta_\mathsf{R}$ in Prop. 3.2.2.

**3.2.4  Proposition**   *Every proof of* $\begin{bmatrix} \Psi \\ \Gamma \end{bmatrix} \vdash \begin{bmatrix} G, \Xi \\ \Lambda \end{bmatrix}$ *has shape*

$$
\begin{bmatrix} \Psi_1 \\ \Gamma_1 \end{bmatrix} \vdash \begin{bmatrix} \Xi \\ \Lambda_1 \end{bmatrix} \quad \cdots \quad \begin{bmatrix} \Psi_h \\ \Gamma_h \end{bmatrix} \vdash \begin{bmatrix} \Xi \\ \Lambda_h \end{bmatrix} \qquad or \qquad \begin{matrix} \circ \\ \mathsf{G_R}\!\uparrow \\ \begin{bmatrix} \Psi \\ \Gamma \end{bmatrix} \vdash \begin{bmatrix} \forall \vec{x}.\top, \Xi \\ \Lambda \end{bmatrix} \end{matrix} \; .
$$

$$
\mathsf{G_R} \qquad \qquad \mathsf{G_R} \\
\begin{bmatrix} \Psi \\ \Gamma \end{bmatrix} \vdash \begin{bmatrix} \forall \vec{x}.(\delta_1 \,\&\, \cdots \,\&\, \delta_h), \Xi \\ \Lambda \end{bmatrix}
$$

**Proof**   By reasoning bottom-up, each application of an inference rule is compulsory.   □

$\mathsf{G_R}$ defines the behaviour of goals when they appear at the right of $\vdash$. They generate as many branches in the computation as there are clauses in the conjunction. When $\mathsf{G_R}$ is applied to a $\forall \vec{x}.\top$, it just *terminates* a (thread of a) computation.

**3.2.5  Definition**   A *G-state sequent* is a state sequent of the kind $\begin{bmatrix} \Psi \\ \Gamma \end{bmatrix} \vdash \begin{bmatrix} \; \\ \Lambda \end{bmatrix}$, where all formulae in $\Psi$ and $\Gamma$ are goals.

By 3.1.2 and 3.2.4 we can always reduce provability of a Forum formula (therefore of a FOLL's one, by 2.11) to provability of some G-state sequents. Moreover, we can always reduce provability of a given formula to the provability of exactly one G-state sequent by employing the double negation equivalence $G \equiv (G \multimap \bot) \multimap \bot$: this last formula is a clause.

## 3.3    Deriving in the Left Context

Let us now turn our attention to left rules and the behaviour of goals and clauses when they appear at the left of $\vdash$, as left focused formulae.

G-state sequents embody a natural notion of state for our computations. To proceed computing from a G-state sequent, clauses from its program must be applied to its atomic context. Left rules come into play: application of clauses is mainly accomplished by $\multimap_{\mathsf{L}}$ and $\Rightarrow_{\mathsf{L}}$ rules. Rules $\invamp_{\mathsf{L}}$ and $\bot_{\mathsf{L}}$ have a role in the applicability of clauses. Rule $\multimap_{\mathsf{L}}$ is also responsible for some non-deterministic choices about the splitting of state into multiple substates. Clauses are applicable to the atomic context $\Lambda$ whenever their heads match a submultiset of $\Lambda$'s atoms. Let us focus first on this matching aspect.

**3.3.1  Definition**   Let $\mathsf{h}$ be the following *head matching* inference rule, where $k \geqslant 0$:

$$
\mathsf{h} \; \frac{}{\begin{bmatrix} \Psi \end{bmatrix} a_1 \invamp \cdots \invamp a_k \vdash \begin{bmatrix} \; \\ a_1, \ldots, a_k \end{bmatrix}} \; .
$$

Fig. 5 shows how $\mathsf{h}$ corresponds to Forum inference rules. The same considerations made above about the associativity of $\&$ hold here for $\invamp$.

**3.3.2  Proposition**   *If the sequent* $\begin{bmatrix} \Psi \\ \Gamma \end{bmatrix} a_1 \invamp \cdots \invamp a_k \vdash \begin{bmatrix} \; \\ \Lambda \end{bmatrix}$ *is provable, then* $\Gamma$ *is empty,* $\Lambda = \{a_1, \ldots, a_k\}_+$ *and the only proof is*

$$
\begin{matrix} \circ \\ r\!\uparrow \\ \begin{bmatrix} \Psi \end{bmatrix} a_1 \invamp \cdots \invamp a_k \vdash \begin{bmatrix} \; \\ a_1, \ldots, a_k \end{bmatrix} \end{matrix}
$$

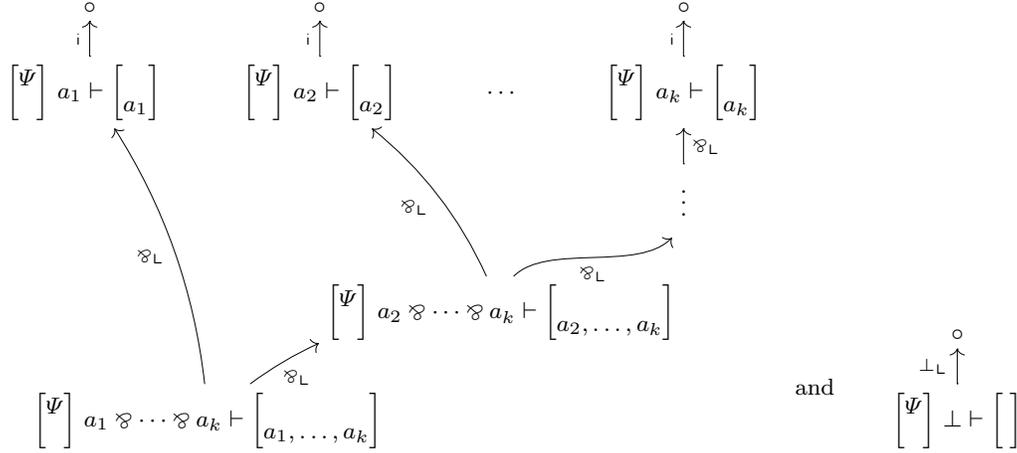

**Fig. 5** *Head matching inference rule* h *when* $k > 0$ *and* $k = 0$

**Proof** Consider Fig. 5: from the root to the leaves, all applications of inference rules are compelled by the left focused formula. Identity axioms force empty left linear contexts. By reading from the leaves to the root, $\mathcal{B}_{\mathsf{L}}$ rules then constrain the conclusion. The case $k = 0$ is trivial. □

**3.3.3 Definition** Let $\delta_{\mathsf{L}}$ be the following *clause reduction left* inference rule, shown in Fig. 6 in terms of h and Forum rules:

$$\delta_{\mathsf{L}} \frac{\begin{bmatrix} \Psi \end{bmatrix} \vdash \begin{bmatrix} G_1 \end{bmatrix} \quad \cdots \quad \begin{bmatrix} \Psi \end{bmatrix} \vdash \begin{bmatrix} G_{k''} \end{bmatrix} \quad \begin{bmatrix} \Psi \\ \Gamma_1 \end{bmatrix} \vdash \begin{bmatrix} H_1 \\ \Lambda_1 \end{bmatrix} \quad \cdots \quad \begin{bmatrix} \Psi \\ \Gamma_{k'} \end{bmatrix} \vdash \begin{bmatrix} H_{k'} \\ \Lambda_{k'} \end{bmatrix}}{\begin{bmatrix} \Psi \\ \Gamma \end{bmatrix} \delta \vdash \begin{bmatrix} \Lambda \end{bmatrix}},$$

where $\delta = G_1 \Rightarrow \cdots \Rightarrow G_{k''} \Rightarrow H_1 \multimap \cdots \multimap H_{k'} \multimap a_1 \mathcal{B} \cdots \mathcal{B} a_k$, and $k, k', k'' \geqslant 0$, and where $\Gamma_1 \uplus \cdots \uplus \Gamma_{k'} = \Gamma$ and $\Lambda_1 \uplus \cdots \uplus \Lambda_{k'} \uplus \{a_1, \ldots, a_k\}_+ = \Lambda$.

Let us go through $\delta_{\mathsf{L}}$ step by step; please note that it contains left rules only. This is a phase in a derivation in which a left focused clause is reduced and used in a rewriting.

1) All classical premises of $\delta$ are evaluated in classical context $\Psi$.

2) All linear premises of $\delta$ are evaluated in classical context $\Psi$ and in linear contexts which are nondeterministically obtained as indicated. Every linear premise gets a part of each piece of linear context: the left one is completely split among the premises; the atomic one is also split except for atoms which have to match the head of the selected clause; the right one is empty. In the figure some relations among contexts are noted for convenience.

3) The head of $\delta$ is matched against the residual atomic context in an h rule.

As an outcome of the reduction of the left focused clause, we have a multiset of premises which will be further reduced by as many $\mathsf{G_R}$ rules. They, in turn, will produce G-state sequents. The most degenerate instances of $\delta_{\mathsf{L}}$ have no premises. Special cases where there are no classical or linear premises are easily inferable from the general scheme provided. Thanks to uniform provability, all non-determinism in searching for Forum proofs resides in left rules. Much of it can be concentrated into a decision rule, but one should notice that $\delta_{\mathsf{L}}$ is also non-deterministic in the splitting of the linear contexts.

**3.3.4 Definition** Let d be the *decision* inference rule, defined by the following two (non-mutually

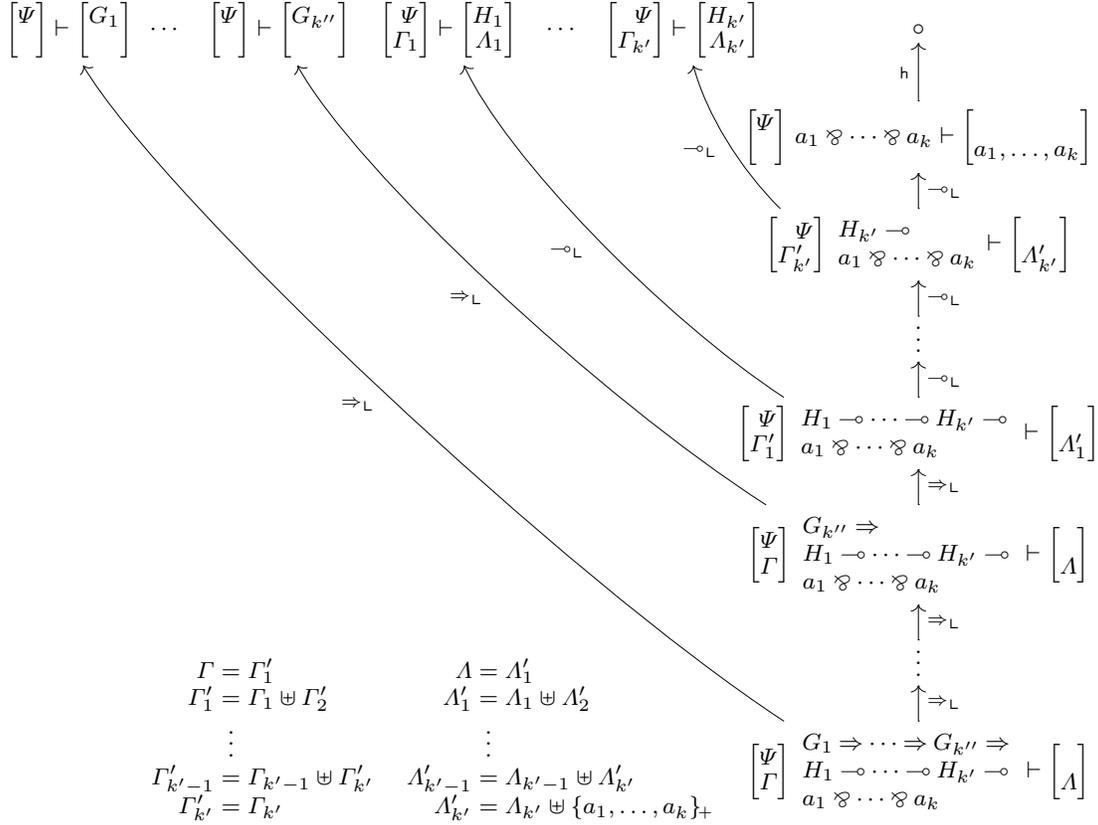

**Fig. 6**  *Clause reduction left inference rule* $\delta_{\mathsf{L}}$

exclusive) cases, and shown in Fig. 7 in terms of Forum rules:

$$\mathsf{d} \frac{\begin{bmatrix} \Psi \\ \Gamma \end{bmatrix} \delta_l \sigma \vdash \begin{bmatrix} \Lambda \end{bmatrix}}{\begin{bmatrix} \Psi \\ \Gamma, G \end{bmatrix} \vdash \begin{bmatrix} \Lambda \end{bmatrix}} \qquad \text{or} \qquad \mathsf{d} \frac{\begin{bmatrix} \Psi, G \\ \Gamma \end{bmatrix} \delta_l \sigma \vdash \begin{bmatrix} \Lambda \end{bmatrix}}{\begin{bmatrix} \Psi, G \\ \Gamma \end{bmatrix} \vdash \begin{bmatrix} \Lambda \end{bmatrix}},$$

where the conclusions are G-state sequents. In the first case $\Gamma, G$ is the *selected context*, in the second it is $\Psi, G$. Goal $G = \forall \vec{x}.(\delta_1 \ \& \ \dots \ \& \ \delta_h)$, where $h > 0$ and $\vec{x}$ can be empty, is the *selected goal*; $\delta_l \sigma$ is the *selected clause*, $1 \leq l \leq h$, and $\sigma$ is a substitution whose domain is $\vec{x}$.

**3.3.5  Proposition**  *All proofs of* $\begin{bmatrix} \Psi \\ \Gamma \end{bmatrix} \vdash \begin{bmatrix} \Lambda \end{bmatrix}$ *have shape*

$$\begin{bmatrix} \Psi \\ \Gamma' \end{bmatrix} \delta \vdash \begin{bmatrix} \Lambda \end{bmatrix},$$

$$\mathsf{d} \uparrow$$

$$\begin{bmatrix} \Psi \\ \Gamma \end{bmatrix} \vdash \begin{bmatrix} \Lambda \end{bmatrix}$$

*for some* $\Gamma'$ *and the inference rule above* $\mathsf{d}$ *is* $\delta_{\mathsf{L}}$.

**Proof**  By reasoning bottom-up, each application of an inference rule is compulsory.  □

**Fig. 7** *Decision inference rule* d *in its two possibilities*

**3.3.6 Remark** There are no proofs for the sequent $\begin{bmatrix} \Psi \\ \Gamma \end{bmatrix} \forall \vec{x}.\top \vdash \begin{bmatrix} \Lambda \end{bmatrix}$.

Each application of d involves the following choices:

1) which goal $G = \forall \vec{x}.(\delta_1 \,\&\, \ldots \,\&\, \delta_h)$ to select, and in which context; it could be the case that a certain $G$ appears both in the classical and in the linear program;

2) which substitution $\sigma$ to apply;

3) which clause $\delta_l \sigma$ to select among $\delta_1 \sigma \,\&\, \ldots \,\&\, \delta_h \sigma$.

We can build on d and $\delta_L$ an inference rule which reduces goals in the program.

**3.3.7 Definition** Let $\mathsf{G_L}$ be the *goal reduction left* inference rule, defined in the following two (non-mutually exclusive) cases and shown in Fig. 8 in terms of d and $\delta_L$ rules:

$$\mathsf{G_L} \frac{\begin{bmatrix} \Psi \end{bmatrix} \vdash \begin{bmatrix} G_1 \end{bmatrix} \;\ldots\; \begin{bmatrix} \Psi \end{bmatrix} \vdash \begin{bmatrix} G_{k''} \end{bmatrix} \quad \begin{bmatrix} \Psi \\ \Gamma_1 \end{bmatrix} \vdash \begin{bmatrix} H_1 \\ \Lambda_1 \end{bmatrix} \;\ldots\; \begin{bmatrix} \Psi \\ \Gamma_{k'} \end{bmatrix} \vdash \begin{bmatrix} H_{k'} \\ \Lambda_{k'} \end{bmatrix}}{\begin{bmatrix} \Psi \\ \Gamma, G \end{bmatrix} \vdash \begin{bmatrix} \Lambda \end{bmatrix}} \quad \text{or}$$

$$\mathsf{G_L} \frac{\begin{bmatrix} \Psi, G \end{bmatrix} \vdash \begin{bmatrix} G_1 \end{bmatrix} \;\ldots\; \begin{bmatrix} \Psi, G \end{bmatrix} \vdash \begin{bmatrix} G_{k''} \end{bmatrix} \quad \begin{bmatrix} \Psi, G \\ \Gamma_1 \end{bmatrix} \vdash \begin{bmatrix} H_1 \\ \Lambda_1 \end{bmatrix} \;\ldots\; \begin{bmatrix} \Psi, G \\ \Gamma_{k'} \end{bmatrix} \vdash \begin{bmatrix} H_{k'} \\ \Lambda_{k'} \end{bmatrix}}{\begin{bmatrix} \Psi, G \\ \Gamma \end{bmatrix} \vdash \begin{bmatrix} \Lambda \end{bmatrix}},$$

where $G = \forall \vec{x}.(\delta_1 \,\&\, \ldots \,\&\, \delta_h)$, $\vec{x}$ can be empty, $\delta_l \sigma = G_1 \Rightarrow \cdots \Rightarrow G_{k''} \Rightarrow H_1 \multimap \cdots \multimap H_{k'} \multimap a_1 \,\rotatebox[origin=c]{180}{\&}\, \cdots \,\rotatebox[origin=c]{180}{\&}\, a_k$, for $1 \leq l \leq h$ and $k, k', k'' \geqslant 0$, and where $\Gamma_1 \uplus \cdots \uplus \Gamma_{k'} = \Gamma$ and $\Lambda_1 \uplus \cdots \uplus \Lambda_{k'} \uplus \{a_1, \ldots, a_k\}_+ = \Lambda$.

## 3.4 The System

In the previous section we built two big inference rules, a left one and a right one. Their definition might look complex, but it is in fact rather straightforward once one knows the (operational) meaning of linear logic connectives. One should notice that if the language of formulae were not restricted to goals, such an enterprise would really be cumbersome and complex, and probably pointless. Our point is, instead, that goals in Forum actually define sort of a generalised connective.

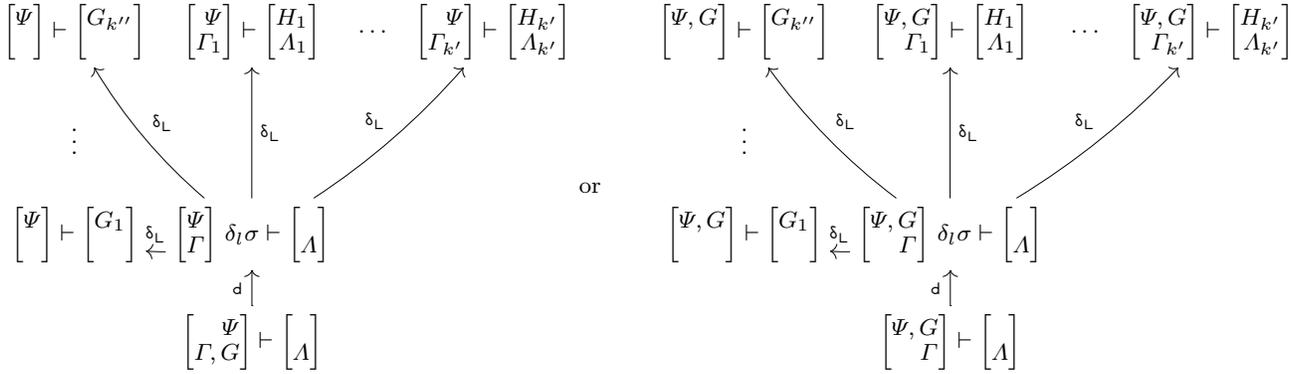

**Fig. 8**  *Goal reduction left inference rule* $\mathsf{G_L}$ *in its two possibilities*

**3.4.1 Definition**  Let G-Forum be the formal system whose sequents are G-state sequents or sequents of the form $\begin{bmatrix} \Psi \\ \Gamma \end{bmatrix} \vdash \begin{bmatrix} G \\ \Lambda \end{bmatrix}$, where $\Psi$ and $\Gamma$ contain goals, and whose inference rules are $\mathsf{G_L}$ and $\mathsf{G_R}$.

An important technical feature of a sequent system is that the rules allow for a cut elimination theorem. We know already cut elimination from linear logic, but that theorem is proved inside the sequent system where all the usual connectives are defined. Even if we know that cut is admissible in our system, in order to test its internal harmony we have to provide a cut elimination proof that does not appeal to the underlying 'small' connectives.

# 4    Cut Elimination

Let us firstly define two natural cut rules for G-Forum.

**4.1    Definition**  The following inference rules $\bowtie_\mathsf{L}$ and $\bowtie_\mathsf{C}$ are respectively called *cut linear* and *cut classical*:

$$\bowtie_\mathsf{L} \frac{\begin{bmatrix} \Psi \\ \Gamma \end{bmatrix} \vdash \begin{bmatrix} G \\ \Lambda \end{bmatrix} \quad \begin{bmatrix} \Psi' \\ G, \Gamma' \end{bmatrix} \vdash \begin{bmatrix} \Xi' \\ \Lambda' \end{bmatrix}}{\begin{bmatrix} \Psi, \Psi' \\ \Gamma, \Gamma' \end{bmatrix} \vdash \begin{bmatrix} \Xi' \\ \Lambda, \Lambda' \end{bmatrix}} \quad \text{and} \quad \bowtie_\mathsf{C} \frac{\begin{bmatrix} \Psi \\ \end{bmatrix} \vdash \begin{bmatrix} G \\ \end{bmatrix} \quad \begin{bmatrix} G, \Psi' \\ \Gamma' \end{bmatrix} \vdash \begin{bmatrix} \Xi' \\ \Lambda' \end{bmatrix}}{\begin{bmatrix} \Psi, \Psi' \\ \Gamma' \end{bmatrix} \vdash \begin{bmatrix} \Xi' \\ \Lambda' \end{bmatrix}} .$$

$\Xi'$ is either empty or a singleton. In both rules $G$ is called the *principal formula*. System G-Forum$^{\bowtie_\mathsf{L}, \bowtie_\mathsf{C}}$ is G-Forum where $\bowtie_\mathsf{L}$, $\bowtie_\mathsf{C}$ are allowed in proofs.

In the sequel we prove that every proof in G-Forum$^{\bowtie_\mathsf{L}, \bowtie_\mathsf{C}}$ can be transformed into an equivalent proof in G-Forum. The proof of the cut elimination theorem follows a traditional argument in which one deals with contraction by a generalised cut rule that cuts on several copies of the same principal formula (see for example [4]). We use the cut classical rule, in a certain generalisation $\bowtie'_\mathsf{C}$ together with a contraction rule $>$, to make G-Forum$^{\bowtie_\mathsf{L}, \bowtie_\mathsf{C}}$ more general, and then we prove cut elimination on this more general system. The core of the proof is the elimination of the $\bowtie_\mathsf{L}$ rule. Actually, the design of the rules $\mathsf{G_L}$ and $\mathsf{G_R}$, and the crucial decisions about the exact definition of goals, all come from a careful analysis of what is needed in this part of the cut elimination argument.

We first introduce a contraction rule and a cut classical rule in a more general form as follows.

**4.2**   **Definition**   The following inference rule $>$ is called *contraction*:

$$> \frac{\left[\begin{smallmatrix} \Psi, G, G \\ \Gamma \end{smallmatrix}\right] \vdash \left[\begin{smallmatrix} \Xi \\ \Lambda \end{smallmatrix}\right]}{\left[\begin{smallmatrix} \Psi, G \\ \Gamma \end{smallmatrix}\right] \vdash \left[\begin{smallmatrix} \Xi \\ \Lambda \end{smallmatrix}\right]}.$$

G-Forum$^{>,\bowtie_{\mathsf{L}},\bowtie_{\mathsf{c}}}$, G-Forum$^{>,\bowtie_{\mathsf{L}}}$ and G-Forum$^{>}$, stand for G-Forum where, in addition to $\mathsf{G_L}$ and $\mathsf{G_R}$, rules in the superscript are allowed.

**4.3**   **Definition**   We write $lG$ to indicate the multiset $\{G, \ldots, G\}_+$ where $G$ appears $l$ times; analogously, given the multiset of formulae $\Psi$, we write $l\Psi$ to indicate $\underbrace{\Psi \uplus \cdots \uplus \Psi}_{l \text{ times}}$; when $l = 0$ both $lG$ and $l\Psi$ stand for $\varnothing_+$.

**4.4**   **Definition**   The *generalized cut classical* rule $\bowtie'_{\mathsf{c}}$ is so defined:

$$\bowtie'_{\mathsf{c}} \frac{\left[\begin{smallmatrix} \Psi \end{smallmatrix}\right] \vdash \left[\begin{smallmatrix} G \end{smallmatrix}\right] \qquad \left[\begin{smallmatrix} lG, \Psi' \\ \Gamma' \end{smallmatrix}\right] \vdash \left[\begin{smallmatrix} \Xi' \\ \Lambda' \end{smallmatrix}\right]}{\left[\begin{smallmatrix} \Psi, \Psi' \\ \Gamma' \end{smallmatrix}\right] \vdash \left[\begin{smallmatrix} \Xi' \\ \Lambda' \end{smallmatrix}\right]},$$

where $l \geqslant 0$. Again, $G$ is called the *principal formula*.

We will prove that proofs in G-Forum$^{>,\bowtie_{\mathsf{L}},\bowtie'_{\mathsf{c}}}$ can be transformed into proofs in G-Forum$^{>,\bowtie_{\mathsf{L}}}$, which in turn can be transformed into proofs in G-Forum$^{>}$, which can be transformed into proofs in G-Forum, where every transformation preserves the conclusion:

$$\text{G-Forum}^{>,\bowtie_{\mathsf{L}},\bowtie'_{\mathsf{c}}} \to \text{G-Forum}^{>,\bowtie_{\mathsf{L}}} \to \text{G-Forum}^{>} \to \text{G-Forum}.$$

The first step and the third one are technically similar and eminently 'structural' in nature, meaning that they deal with bookkeeping of resources. The middle step is a typical cut elimination theorem which tests the symmetries of the system.

The cut elimination theorem relies on an induction measure that we call *cut-rank*, which is essentially a measure of the syntactical complexity of the principal formula.

**4.5**   **Definition**   The *cut-rank* $\mathsf{cr}(G)$ of goal $G$ is a natural number inductively defined as follows:

$$\mathsf{cr}(\forall \vec{x}.\delta_1 \, \& \ldots \& \, \delta_h) = \mathsf{max}\{\mathsf{cr}'(\delta_1), \ldots, \mathsf{cr}'(\delta_h), 0\} + 1, \qquad \text{where } h \geqslant 0;$$
$$\mathsf{cr}'(G_1 \Rightarrow \cdots \Rightarrow G_{k''} \Rightarrow H_1 \multimap \cdots \multimap H_{k'} \multimap a_1 \bindnasrepma \cdots \bindnasrepma a_k) =$$
$$\mathsf{max}\{\mathsf{cr}(G_1), \ldots, \mathsf{cr}(G_{k''}), \mathsf{cr}(H_1), \ldots, \mathsf{cr}(H_{k'}), 0\} + 1, \quad \text{where } k, k', k'' \geqslant 0.$$

The *cut-rank* of an instance of $\bowtie_{\mathsf{L}}$, $\bowtie_{\mathsf{c}}$ or $\bowtie'_{\mathsf{c}}$ rule is the cut-rank of its principal formula. The cut-rank $\mathsf{cr}(\Pi)$ of a proof $\Pi$ is the maximum cut-rank of instances of $\bowtie_{\mathsf{L}}$, $\bowtie_{\mathsf{c}}$ and $\bowtie'_{\mathsf{c}}$ in $\Pi$; if $\Pi$ is cut-free then we define $\mathsf{cr}(\Pi) = 0$.

### 4.1   Elimination of the Generalized Cut Classical Rule

To prove that $\bowtie'_{\mathsf{c}}$ occurrences may be eliminated from a proof, we prove that every instance of $\bowtie'_{\mathsf{c}}$ can be moved upwards in the proof, along the right branch of the rule, until it disappears.

**4.1.1 Lemma**  *In* G-Forum$^{>,\bowtie_{\mathsf{L}},\bowtie'_{\mathsf{C}}}$, *let* $\Pi$ *be the proof*

$$
\begin{array}{ccc}
\overline{\Pi'} & & \overline{\Pi''} \\[4pt]
\begin{bmatrix}\Psi\end{bmatrix} \vdash \begin{bmatrix}G\end{bmatrix} & & \begin{bmatrix}lG,\Psi' \\ \Gamma'\end{bmatrix} \vdash \begin{bmatrix}\Xi' \\ \Lambda'\end{bmatrix} \\[12pt]
\llap{{\scriptstyle\bowtie'_{\mathsf{C}}}}\nwarrow & & \nearrow\rlap{{\scriptstyle\bowtie'_{\mathsf{C}}}} \\[6pt]
& \begin{bmatrix}\Psi,\Psi' \\ \Gamma'\end{bmatrix} \vdash \begin{bmatrix}\Xi' \\ \Lambda'\end{bmatrix} &
\end{array}\quad,
$$

*where* $l \geqslant 0$ *and no instance of* $\bowtie'_{\mathsf{C}}$ *appears in* $\Pi'$ *and* $\Pi''$. *Then, there exists a proof* $\bar{\Pi}$ *in* G-Forum$^{>,\bowtie_{\mathsf{L}}}$, *whose conclusion is the same as that of* $\Pi$, *and* $\mathsf{cr}(\bar{\Pi}) \leq \mathsf{cr}(\Pi)$.

**Proof**  We will prove the lemma by induction on the depth of $\Pi''$.

**Basis Cases**

The following two cases are the only possible ones:

1)  $\mathsf{G_L}$. If

$$
\Pi = \quad
\begin{array}{ccc}
\overline{\Pi'} & & \overset{\circ}{\underset{\mathsf{G_L}}{\uparrow}} \\[4pt]
\begin{bmatrix}\Psi\end{bmatrix} \vdash \begin{bmatrix}G\end{bmatrix} & & \begin{bmatrix}lG,\Psi' \\ \Gamma'\end{bmatrix} \vdash \begin{bmatrix}\Lambda'\end{bmatrix} \\[12pt]
\llap{{\scriptstyle\bowtie'_{\mathsf{C}}}}\nwarrow & & \nearrow\rlap{{\scriptstyle\bowtie'_{\mathsf{C}}}} \\[6pt]
& \begin{bmatrix}\Psi,\Psi' \\ \Gamma'\end{bmatrix} \vdash \begin{bmatrix}\Lambda'\end{bmatrix} &
\end{array}\quad,
$$

then two cases are possible. If $G$ is not the selected goal in the $\mathsf{G_L}$ instance, or it is selected and either $G \in \Psi'$ or $G \in \Gamma'$, then take:

$$
\bar{\Pi} = \quad
\begin{array}{c}
\overset{\circ}{\underset{\mathsf{G_L}}{\uparrow}} \\[4pt]
\begin{bmatrix}\Psi,\Psi' \\ \Gamma'\end{bmatrix} \vdash \begin{bmatrix}\Lambda'\end{bmatrix}
\end{array}\quad.
$$

Otherwise, if $G$ is the selected goal in the $\mathsf{G_L}$ instance and $G$ does not appear neither in $\Psi'$ nor in $\Gamma'$, it must be $l > 0$ and $\Gamma' = \varnothing_+$; in this case take

$$
\bar{\Pi} = \quad
\begin{array}{ccc}
\overline{\Pi'} & & \overset{\circ}{\underset{\mathsf{G_L}}{\uparrow}} \\[4pt]
\begin{bmatrix}\Psi\end{bmatrix} \vdash \begin{bmatrix}G\end{bmatrix} & & \begin{bmatrix}\Psi' \\ G\end{bmatrix} \vdash \begin{bmatrix}\Lambda'\end{bmatrix} \\[12pt]
\llap{{\scriptstyle\bowtie_{\mathsf{L}}}}\nwarrow & & \nearrow\rlap{{\scriptstyle\bowtie_{\mathsf{L}}}} \\[6pt]
& \begin{bmatrix}\Psi,\Psi'\end{bmatrix} \vdash \begin{bmatrix}\Lambda'\end{bmatrix} &
\end{array}\quad.
$$

2)  $\mathsf{G_R}$. If

$$
\Pi = \quad
\begin{array}{ccc}
\overline{\Pi'} & & \overset{\circ}{\underset{\mathsf{G_R}}{\uparrow}} \\[4pt]
\begin{bmatrix}\Psi\end{bmatrix} \vdash \begin{bmatrix}G\end{bmatrix} & & \begin{bmatrix}lG,\Psi' \\ \Gamma'\end{bmatrix} \vdash \begin{bmatrix}\forall\vec{x}.\top \\ \Lambda'\end{bmatrix} \\[12pt]
\llap{{\scriptstyle\bowtie'_{\mathsf{C}}}}\nwarrow & & \nearrow\rlap{{\scriptstyle\bowtie'_{\mathsf{C}}}} \\[6pt]
& \begin{bmatrix}\Psi,\Psi' \\ \Gamma'\end{bmatrix} \vdash \begin{bmatrix}\forall\vec{x}.\top \\ \Lambda'\end{bmatrix} &
\end{array}\quad,
$$

then take:

$$\bar{\Pi} = \begin{array}{c} \circ \\ {}^{\mathsf{G_R}}\!\uparrow \\ \left[\begin{smallmatrix}\Psi, \Psi' \\ \Gamma'\end{smallmatrix}\right] \vdash \left[\begin{smallmatrix}\forall \vec{x}. \top \\ \Lambda'\end{smallmatrix}\right] \end{array} \quad .$$

In all cases it holds $\mathsf{cr}(\bar{\Pi}) \leq \mathsf{cr}(\Pi)$.

**Inductive Cases**

3)    $\mathsf{G_L}$. Let $\Pi$ be

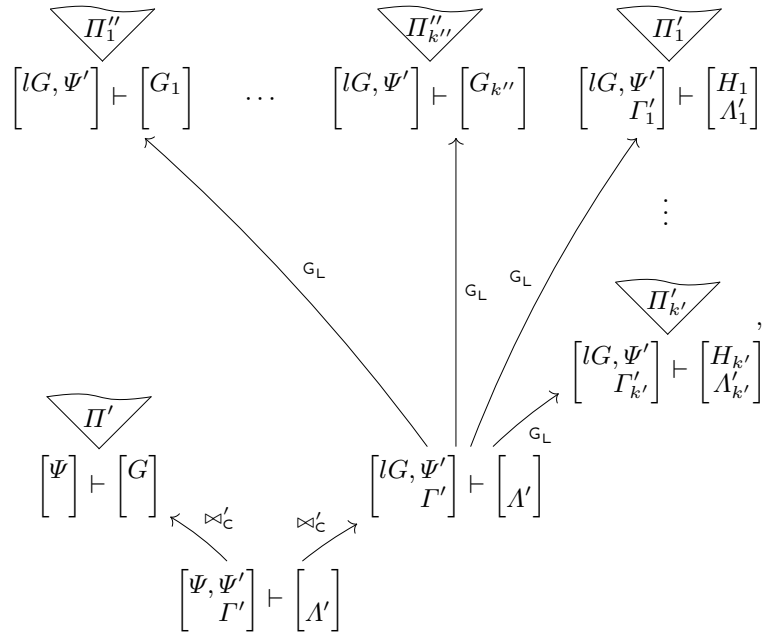

where $k' + k'' > 0$. By the induction hypothesis, for $1 \leq j \leq k''$ and $1 \leq i \leq k'$, there are $\bowtie_{\mathsf{C}}'$-free proofs $\bar{\Pi}_j''$ and $\bar{\Pi}_i$, corresponding respectively to

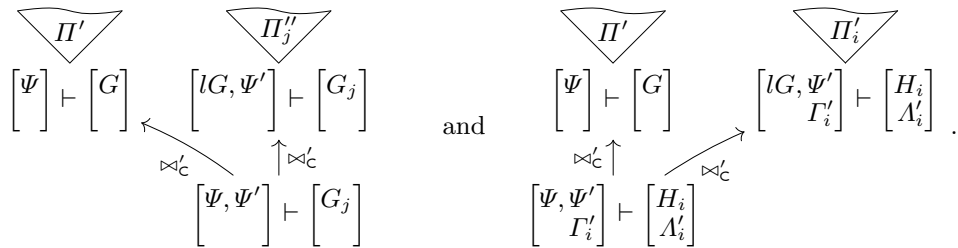

Two cases are possible. If $G$ is not the selected goal in the $\mathsf{G_L}$ instance, or if it is selected and either

$G \in \Psi'$ or $G \in \Gamma'$, then take $\bar{\Pi}$ as:

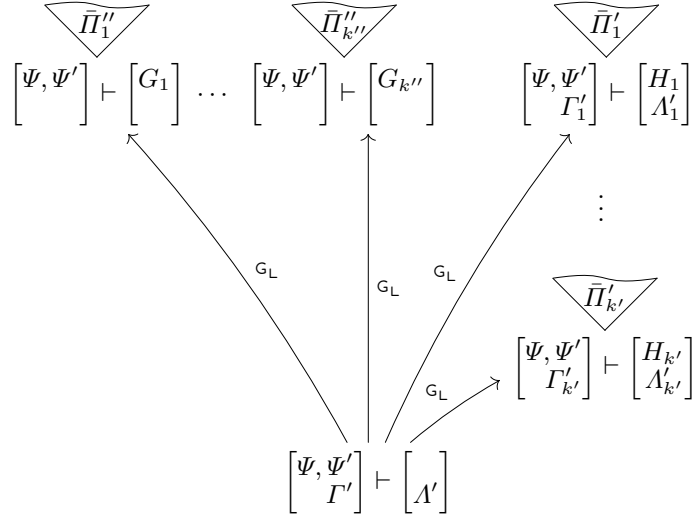

Otherwise, if $G$ is selected and $G$ does not appear neither in $\Psi'$ nor in $\Gamma'$, it must be $l > 0$; in this case take $\bar{\Pi}$ as

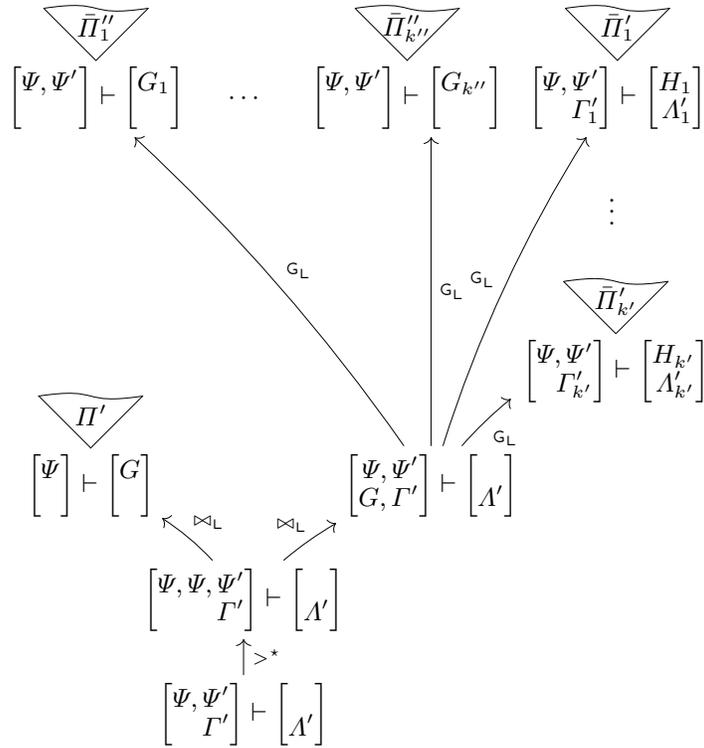

4)   $\mathsf{G_R}$. Let $\Pi$ be

$$
\begin{array}{c}
\overline{\Pi'} \\[2pt]
\left[\Psi\right] \vdash \left[G\right]
\end{array}
\qquad
\begin{array}{c}
\overline{\Pi_1} \\[2pt]
\left[\begin{smallmatrix} lG, \Psi', \mathsf{cp}(\delta_1\rho) \\ \Gamma', \mathsf{lp}(\delta_1\rho) \end{smallmatrix}\right] \vdash \left[\mathsf{hd}(\delta_1\rho), \Lambda'\right]
\end{array}
\;\cdots\;
\begin{array}{c}
\overline{\Pi_h} \\[2pt]
\left[\begin{smallmatrix} lG, \Psi', \mathsf{cp}(\delta_h\rho) \\ \Gamma', \mathsf{lp}(\delta_h\rho) \end{smallmatrix}\right] \vdash \left[\mathsf{hd}(\delta_h\rho), \Lambda'\right]
\end{array}
\qquad ,
$$

$$
\begin{array}{c}
\left[\begin{smallmatrix} lG, \Psi' \\ \Gamma' \end{smallmatrix}\right] \vdash \left[\begin{smallmatrix} G' \\ \Lambda' \end{smallmatrix}\right] \quad\mathsf{G_R}\\[6pt]
\bowtie'_{\mathsf{C}}\\[6pt]
\left[\begin{smallmatrix} \Psi, \Psi' \\ \Gamma' \end{smallmatrix}\right] \vdash \left[\begin{smallmatrix} G' \\ \Lambda' \end{smallmatrix}\right]
\end{array}
$$

where $G' = \forall \vec{x}.(\delta_1 \,\&\, \ldots \,\&\, \delta_h)$, for $h > 0$, and $\rho$ is a renaming substitution over domain $\vec{x}$; we can assume that no variable in the range of $\rho$ is free in $\Psi$. By the induction hypothesis, for $1 \le i \le h$, there are $\bowtie'_{\mathsf{C}}$-free proofs $\bar{\Pi}_i$, corresponding to:

$$
\begin{array}{c}
\overline{\Pi'} \\[2pt]
\left[\Psi\right] \vdash \left[G\right]
\end{array}
\qquad
\begin{array}{c}
\overline{\Pi_i} \\[2pt]
\left[\begin{smallmatrix} lG, \Psi', \mathsf{cp}(\delta_i\rho) \\ \Gamma', \mathsf{lp}(\delta_i\rho) \end{smallmatrix}\right] \vdash \left[\mathsf{hd}(\delta_i\rho), \Lambda'\right]
\end{array}
\qquad .
$$

$$
\begin{array}{c}
\bowtie'_{\mathsf{C}}\\[6pt]
\left[\begin{smallmatrix} \Psi, \Psi', \mathsf{cp}(\delta_i\rho) \\ \Gamma', \mathsf{lp}(\delta_i\rho) \end{smallmatrix}\right] \vdash \left[\mathsf{hd}(\delta_i\rho), \Lambda'\right]
\end{array}
$$

Take

$$
\bar{\Pi} = 
\begin{array}{c}
\overline{\bar{\Pi}_1} \\[2pt]
\left[\begin{smallmatrix} \Psi, \Psi', \mathsf{cp}(\delta_1\rho) \\ \Gamma', \mathsf{lp}(\delta_1\rho) \end{smallmatrix}\right] \vdash \left[\mathsf{hd}(\delta_1\rho), \Lambda'\right]
\end{array}
\;\cdots\;
\begin{array}{c}
\overline{\bar{\Pi}_h} \\[2pt]
\left[\begin{smallmatrix} \Psi, \Psi', \mathsf{cp}(\delta_h\rho) \\ \Gamma', \mathsf{lp}(\delta_h\rho) \end{smallmatrix}\right] \vdash \left[\mathsf{hd}(\delta_h\rho), \Lambda'\right]
\end{array}
\qquad .
$$

$$
\begin{array}{c}
\mathsf{G_R}\\[6pt]
\left[\begin{smallmatrix} \Psi, \Psi' \\ \Gamma' \end{smallmatrix}\right] \vdash \left[\begin{smallmatrix} G' \\ \Lambda' \end{smallmatrix}\right]
\end{array}
$$

5)   $\bowtie_{\mathsf{L}}$. Let $\Pi$ be

$$
\begin{array}{c}
\overline{\Pi'} \\[2pt]
\left[\Psi\right] \vdash \left[G\right]
\end{array}
\qquad
\begin{array}{c}
\overline{\Pi_1} \\[2pt]
\left[\begin{smallmatrix} l_1 G, \Psi'_1 \\ \Gamma'_1 \end{smallmatrix}\right] \vdash \left[\begin{smallmatrix} G' \\ \Lambda'_1 \end{smallmatrix}\right]
\end{array}
\qquad
\begin{array}{c}
\overline{\Pi_h} \\[2pt]
\left[\begin{smallmatrix} l_2 G, \Psi'_2 \\ G', \Gamma'_2 \end{smallmatrix}\right] \vdash \left[\begin{smallmatrix} \Xi' \\ \Lambda'_2 \end{smallmatrix}\right]
\end{array}
\qquad ,
$$

$$
\begin{array}{c}
\bowtie_{\mathsf{L}}\\[6pt]
\left[\begin{smallmatrix} (l_1 + l_2)G, \Psi'_1, \Psi'_2 \\ \Gamma'_1, \Gamma'_2 \end{smallmatrix}\right] \vdash \left[\begin{smallmatrix} \Xi' \\ \Lambda'_1, \Lambda'_2 \end{smallmatrix}\right]\\[6pt]
\bowtie'_{\mathsf{C}}\\[6pt]
\left[\begin{smallmatrix} \Psi, \Psi'_1, \Psi'_2 \\ \Gamma'_1, \Gamma'_2 \end{smallmatrix}\right] \vdash \left[\begin{smallmatrix} \Xi' \\ \Lambda'_1, \Lambda'_2 \end{smallmatrix}\right]
\end{array}
$$

where $l_1, l_2 \geqslant 0$. By the induction hypothesis, consider $\bowtie'_c$-free proofs $\bar{\Pi}_1$ and $\bar{\Pi}_2$ corresponding respectively to

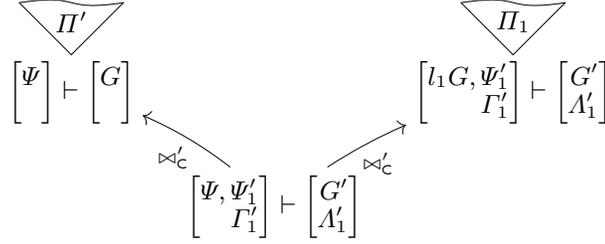

and

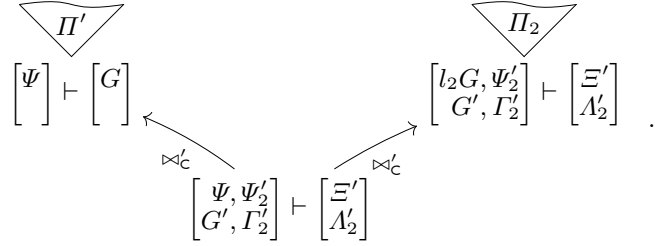

.

Take

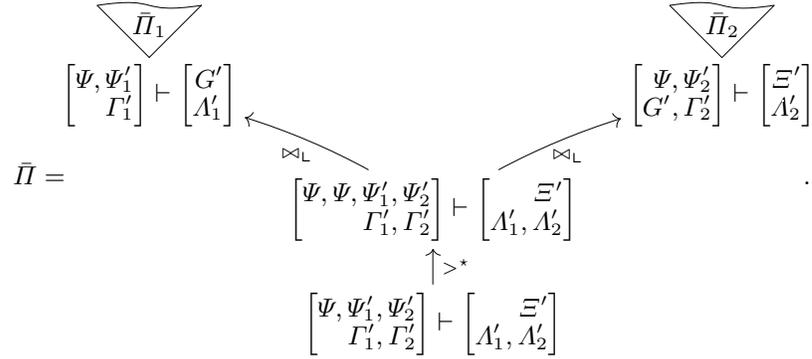

.

6)   $>$. There are two distinct cases. If $\Pi$ is

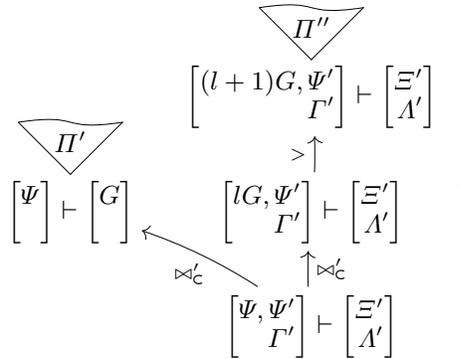

,

where $l > 0$, then, by induction hypothesis, there is a $\bowtie'_c$-free proof $\bar{\Pi}$ corresponding to

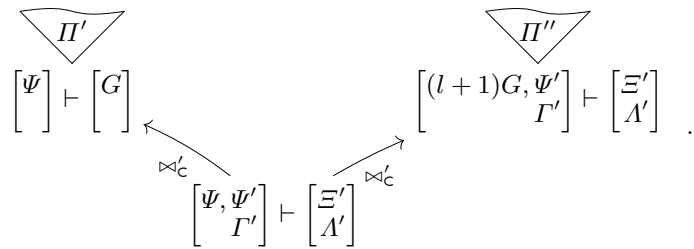

.

If $\Pi$ is

$$
\begin{array}{c}
\overbrace{\phantom{xxx}}^{\Pi''} \\
\left[ \begin{array}{c} lG, G', G', \Psi'' \\ \Gamma' \end{array} \right] \vdash \left[ \begin{array}{c} \Xi' \\ \Lambda' \end{array} \right] \\
{>\uparrow} \\
\left[ \begin{array}{c} lG, G', \Psi'' \\ \Gamma' \end{array} \right] \vdash \left[ \begin{array}{c} \Xi' \\ \Lambda' \end{array} \right] \\
{\uparrow}_{\bowtie'_{\mathsf{c}}} \\
\left[ \begin{array}{c} \Psi, G', \Psi'' \\ \Gamma' \end{array} \right] \vdash \left[ \begin{array}{c} \Xi' \\ \Lambda' \end{array} \right]
\end{array}
$$

with $\Pi'$: $\left[ \Psi \right] \vdash \left[ G \right]$ connected via $\bowtie'_{\mathsf{c}}$ ,

by the induction hypothesis, consider the $\bowtie'_{\mathsf{c}}$-free proof $\bar{\Pi}''$ corresponding to

$$
\begin{array}{c}
\overbrace{\phantom{xxx}}^{\Pi'} \quad\quad\quad\quad \overbrace{\phantom{xxx}}^{\Pi''} \\
\left[ \Psi \right] \vdash \left[ G \right] \quad\quad\quad \left[ \begin{array}{c} lG, G', G', \Psi'' \\ \Gamma' \end{array} \right] \vdash \left[ \begin{array}{c} \Xi' \\ \Lambda' \end{array} \right] \\
{}_{\bowtie'_{\mathsf{c}}}\nwarrow \qquad \nearrow_{\bowtie'_{\mathsf{c}}} \\
\left[ \begin{array}{c} \Psi, G', G', \Psi'' \\ \Gamma' \end{array} \right] \vdash \left[ \begin{array}{c} \Xi' \\ \Lambda' \end{array} \right]
\end{array} \quad .
$$

Take

$$
\bar{\Pi} = \quad
\begin{array}{c}
\overbrace{\phantom{xxx}}^{\bar{\Pi}''} \\
\left[ \begin{array}{c} \Psi, G', G', \Psi'' \\ \Gamma' \end{array} \right] \vdash \left[ \begin{array}{c} \Xi' \\ \Lambda' \end{array} \right] \\
{>\uparrow} \\
\left[ \begin{array}{c} \Psi, G', \Psi'' \\ \Gamma' \end{array} \right] \vdash \left[ \begin{array}{c} \Xi' \\ \Lambda' \end{array} \right]
\end{array} \quad .
$$

In all cases it is easy to check that $\mathsf{cr}(\bar{\Pi}) \leq \mathsf{cr}(\Pi)$. □

**4.1.2 Lemma** *For every proof $\Pi$ in* G-Forum$^{>,\bowtie_{\mathsf{L}},\bowtie'_{\mathsf{c}}}$ *a proof $\Pi'$ exists in* G-Forum$^{>,\bowtie_{\mathsf{L}}}$ *whose conclusion is the same and* $\mathsf{cr}(\Pi') \leq \mathsf{cr}(\Pi)$.

**Proof** If a $\bowtie'_{\mathsf{c}}$ instance appears in the proof $\Pi$, there must be in $\Pi$ a subproof $\Pi'$ which satisfies the hypotheses of Lemma 4.1.1. Replace $\Pi'$ with its associated $\bowtie'_{\mathsf{c}}$-free proof $\bar{\Pi}'$ produced by the lemma. Proceed by induction on the number of $\bowtie'_{\mathsf{c}}$ instances in $\Pi$: at each step the cut-rank of the proof does not increase. □

## 4.2 Elimination of the Cut Linear Rule

In this section we see the core argument of the whole proof.

**4.2.1 Lemma** *If there is in* G-Forum$^{>,\bowtie_{\mathsf{L}}}$ *a proof with conclusion* $\left[ \begin{array}{c} \Psi \\ \Gamma \end{array} \right] \vdash \left[ \begin{array}{c} \Xi \\ \Lambda \end{array} \right]$, *there is also a proof with conclusion* $\left[ \begin{array}{c} \Psi, \Psi' \\ \Gamma \end{array} \right] \vdash \left[ \begin{array}{c} \Xi \\ \Lambda \end{array} \right]$ *and whose cut-rank is the same, for any multiset $\Psi'$.*

**Proof** Add $\Psi'$ to the classical context in the conclusion of the proof and proceed recursively upwards: when a cut-rule is reached, proceed recursively on one of the two proofs ending in its premises; when a rule different from a cut is reached, proceed recursively on all the proofs ending in its premises. The cut-rank is unchanged. □

**4.2.2 Lemma**  *In* G-Forum$^{>,\bowtie_L}$ *let $\Pi$ be the proof*

*such that* $\mathsf{cr}(G) > \mathsf{cr}(\Pi')$ *and* $\mathsf{cr}(G) > \mathsf{cr}(\Pi'')$. *A proof $\bar\Pi$ exists in* G-Forum$^{>,\bowtie_L}$, *whose conclusion is the same as that of $\Pi$, and* $\mathsf{cr}(\bar\Pi) < \mathsf{cr}(\Pi)$.

**Proof**  The proof is by induction on $\mathsf{d}(\Pi') + \mathsf{d}(\Pi'')$.

**Basis Case**

The only basis case occurs when $\Pi$ is

in this case consider

**Inductive Cases**

We consider the bottommost rule instances in $\Pi'$ and $\Pi''$. The bottommost rule in $\Pi'$ cannot be a $\mathsf{G_L}$, so we can make an exhaustive case analysis by considering the following couples 'left bottommost, right bottommost':

1)  $\mathsf{G_R}$, $\mathsf{G_L}$ (where the principal formula is active);

2)  any rule, $\mathsf{G_L}$ (where the principal formula is not active);

3)  any rule, $\mathsf{G_R}$;

4)  any rule, $\bowtie_L$ (principal formula goes to the left);

5)  any rule, $\bowtie_L$ (principal formula goes to the right);

6)  any rule, $>$;

7)  $\bowtie_L$, any rule;

8)  $>$, any rule.

In detail:

1)    $\mathsf{G_R}$, $\mathsf{G_L}$ (where the principal formula is active). Let $\Pi$ be

where $G = \forall \vec{x}.(\delta_1 \,\&\, \cdots \,\&\, \delta_h)$ is the selected goal in the $\mathsf{G_L}$ instance and $h > 0$ (the case $h = 0$ is impossible); the selected clause in $\mathsf{G_L}$ is $\delta_l \sigma = (G_1 \Rightarrow \cdots \Rightarrow G_{k''} \Rightarrow H_1 \multimap \cdots \multimap H_{k'} \multimap a_1 \,\otimes \cdots \otimes\, a_k)\sigma$, where $k, k', k'' \geqslant 0$ and $1 \leq l \leq h$; for some renaming substitution $\rho$ they hold $\Psi_l = \Psi \uplus \{G_1 \rho, \ldots, G_{k''} \rho\}_+$, $\Gamma_l = \Gamma \uplus \{H_1 \rho, \ldots, H_{k'} \rho\}_+$ and $\Lambda_l = \Lambda \uplus \{a_1 \rho, \ldots, a_k \rho\}_+$; $\Gamma' = \Gamma'_1 \uplus \cdots \uplus \Gamma'_{k'}$ and $\Lambda' = \Lambda'_1 \uplus \cdots \uplus \Lambda'_{k'} \uplus \{a_1 \sigma, \ldots, a_k \sigma\}_+$. Let $\tau$ be a substitution whose domain is the range of $\rho$ and such that $\rho \tau = \sigma$. Consider the proof $\bar{\Pi}_l$ obtained from $\Pi_l$ by applying $\tau$ to every formula in every sequent (this operation may of course involve the renaming of some eigenvariable); moreover $\Psi'$ is added to the classical context (Lemma 4.2.1):

$$\overbrace{\begin{bmatrix} G_1\sigma, \ldots, G_{k''}\sigma, \Psi, \Psi' \\ H_1\sigma, \ldots, H_{k'}\sigma, \Gamma \end{bmatrix}}^{\bar{\Pi}_l} \vdash \begin{bmatrix} \Lambda, a_1\sigma, \ldots, a_k\sigma \end{bmatrix}.$$

Of course, $\mathsf{cr}(\bar{\Pi}_l) = \mathsf{cr}(\Pi_l)$. Let $\bar{\Pi}'$ be the following proof:

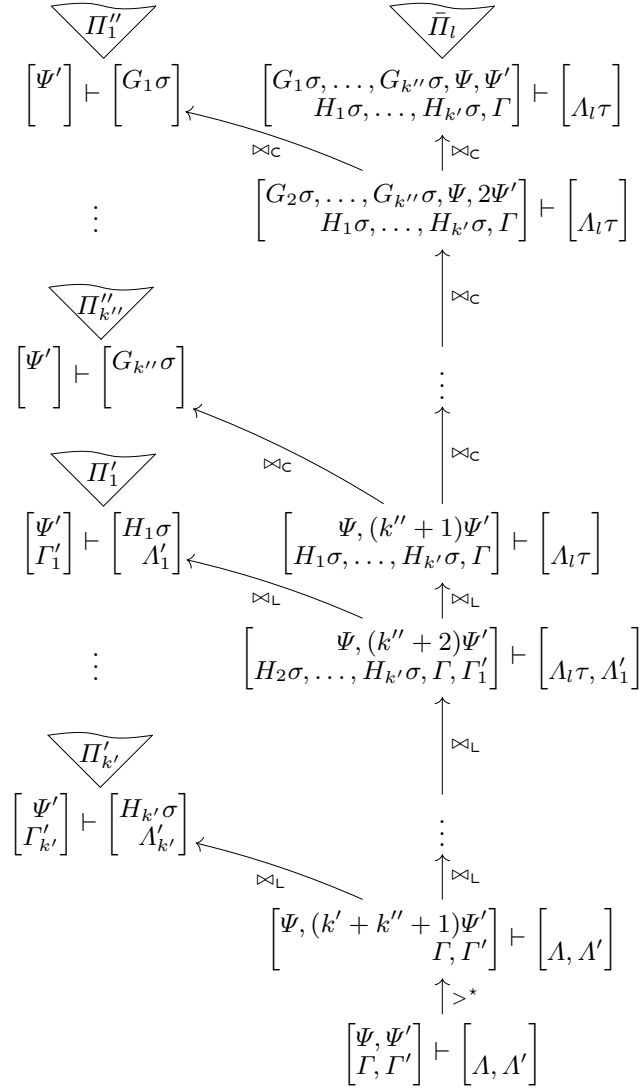

We have

$$\mathsf{cr}(\bar{\Pi}') = \mathsf{max}\{\mathsf{cr}(\bar{\Pi}_l), \mathsf{cr}(\Pi_1''), \ldots, \mathsf{cr}(\Pi_{k''}''), \mathsf{cr}(\Pi_1'), \ldots, \mathsf{cr}(\Pi_{k'}'),$$
$$\mathsf{cr}(G_1\sigma), \ldots, \mathsf{cr}(G_{k''}\sigma), \mathsf{cr}(H_1\sigma), \ldots, \mathsf{cr}(H_{k'}\sigma)\}$$
$$< \mathsf{cr}(G).$$

Apply Lemma 4.1.2 to $\bar{\Pi}'$ and obtain $\bar{\Pi}$: the cut-rank does not increase.

2)    Any rule, $\mathsf{G_L}$ (where the principal formula is not active). Let $\Pi$ be

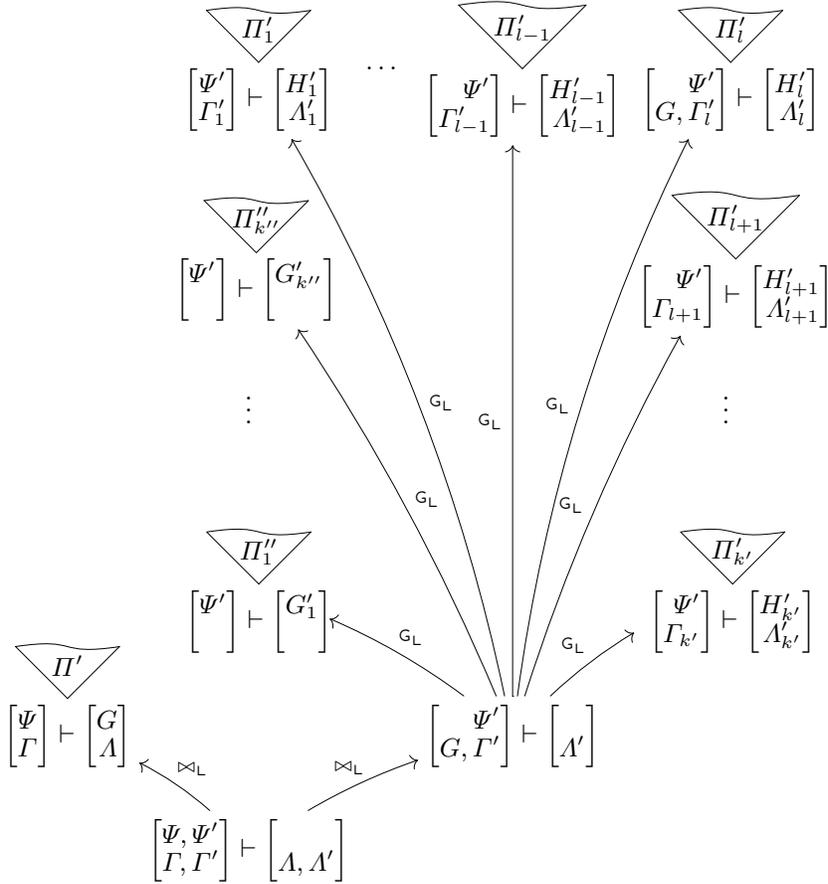

where $k'' \geqslant 0$, goal $G$ is not selected, and then $k' > 0$ and $1 \leq l \leq k'$. Let $\bar{\Pi}_1'', \ldots, \bar{\Pi}_{k''}'', \bar{\Pi}_1', \ldots, \bar{\Pi}_{l-1}',$ $\bar{\Pi}_{l+1}', \ldots, \bar{\Pi}_{k'}'$ respectively be obtained by $\Pi_1'', \ldots, \Pi_{k''}'', \Pi_1', \ldots, \Pi_{l-1}', \Pi_{l+1}', \ldots, \Pi_{k'}'$ by adding $\Psi$ to the classical context in the conclusion (by Lemma 4.2.1). By the induction hypothesis there is a proof $\bar{\Pi}_l'$ with the same conclusion of

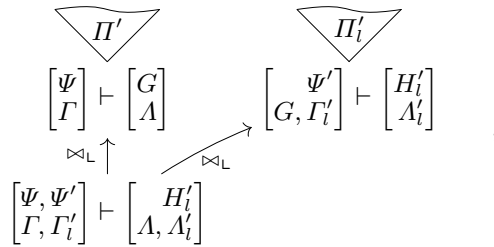

,

and such that $\mathsf{cr}(\bar{\Pi}'_l) < \mathsf{cr}(G)$. Take $\bar{\Pi}$ as

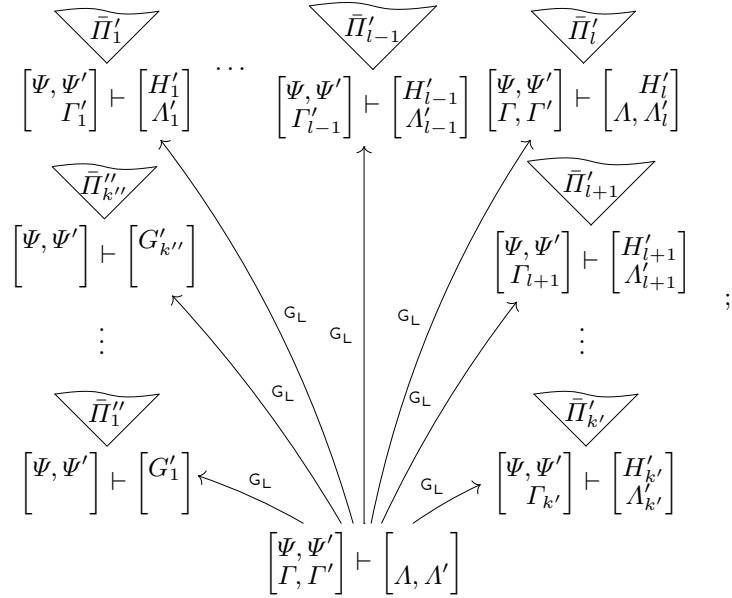

$\mathsf{cr}(\bar{\Pi}) < \mathsf{cr}(\Pi)$ is easily verified.

3) Any rule, $\mathsf{G_R}$. Let $\Pi$ be

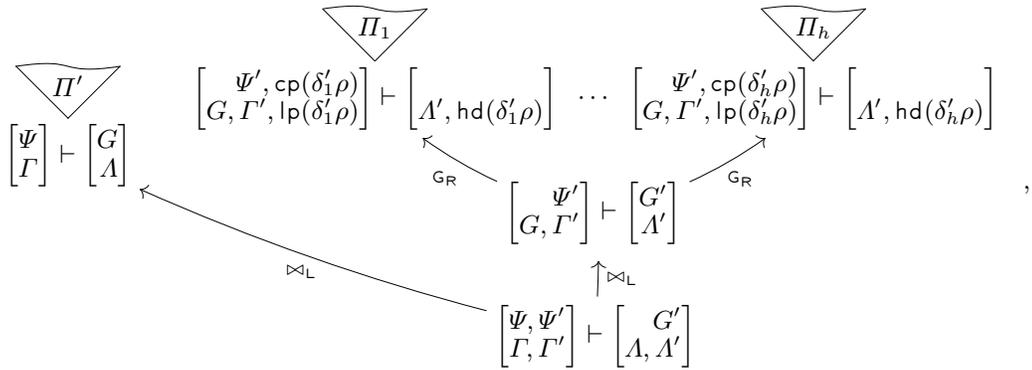

where $G' = \forall \vec{x}.(\delta'_1 \,\&\, \cdots \,\&\, \delta'_h)$ and $h > 0$. Obtain, using the induction hypothesis, proofs $\bar{\Pi}_i$, where $1 \le i \le h$, and such that $\mathsf{cr}(\bar{\Pi}_i) < \mathsf{cr}(G)$, from proofs

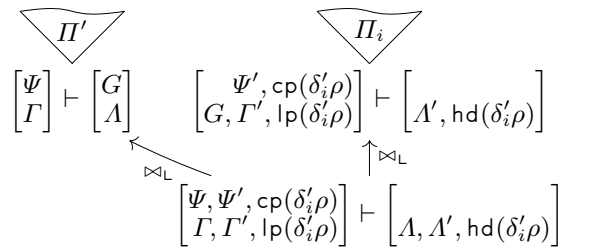

Take $\bar{\Pi}$ as

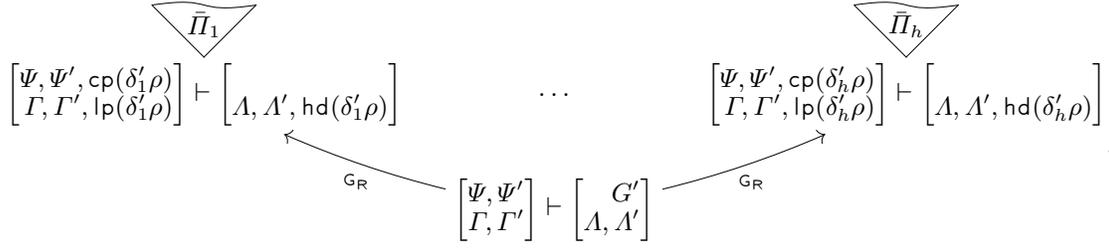

.

When $h = 0$ take $\bar{\Pi}$ as

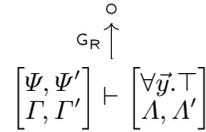

.

4)   Any rule, $\bowtie_\mathsf{L}$ (principal formula goes to the left). Let $\Pi$ be

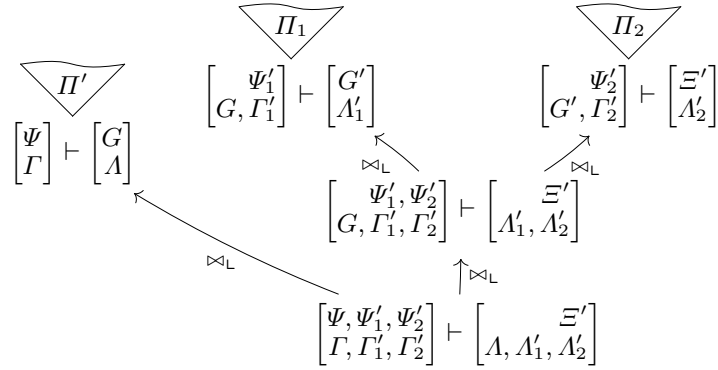

.

Apply the induction hypothesis on

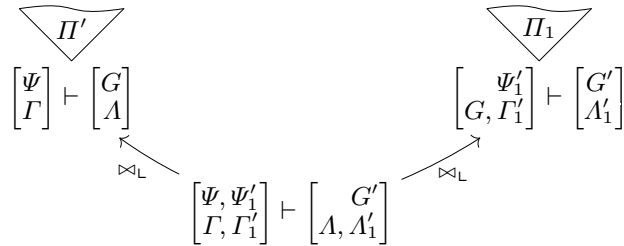

and obtain $\bar{\Pi}_1$. Then take $\bar{\Pi}$ as

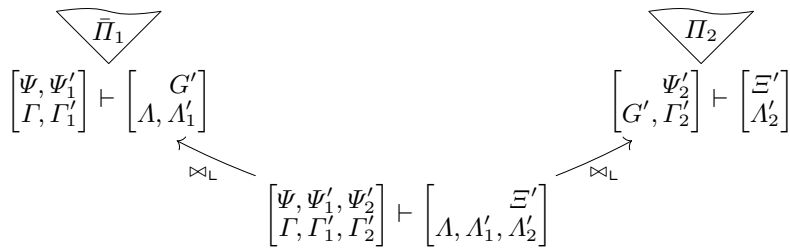

.

5) Any rule, $\bowtie_L$ (principal formula goes to the right). Let $\Pi$ be

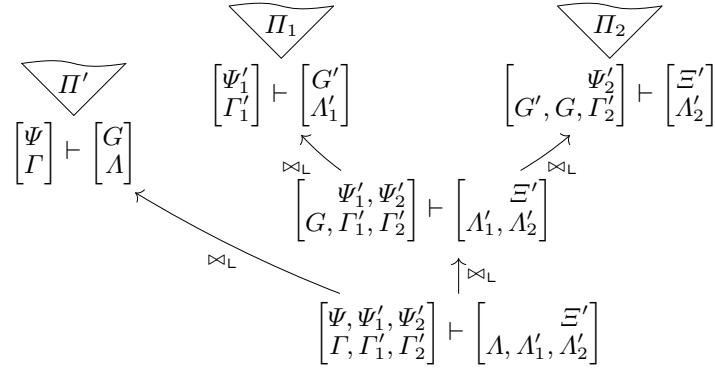

Apply the induction hypothesis on

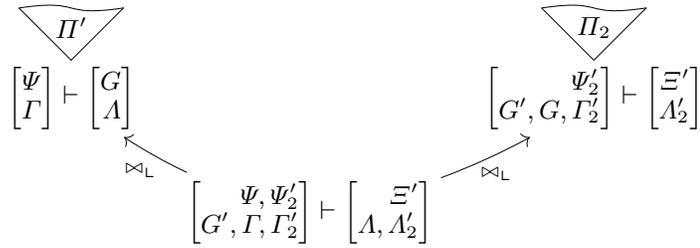

and obtain $\bar{\Pi}_2$. Then take $\bar{\Pi}$ as

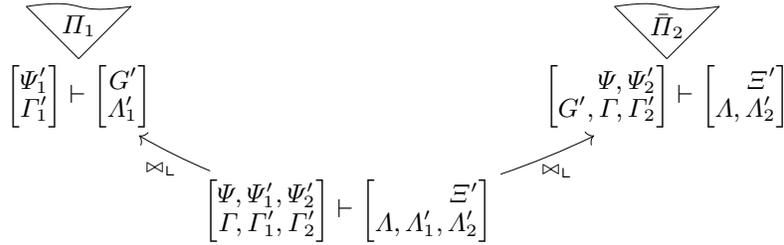

6) Any rule, $>$. Let $\Pi$ be

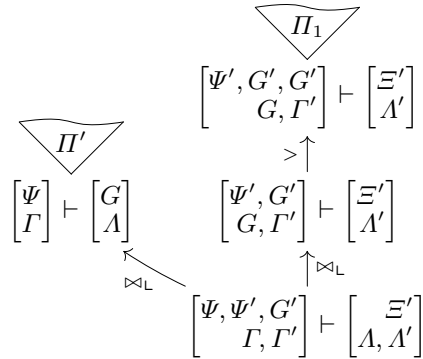

Apply the induction hypothesis on

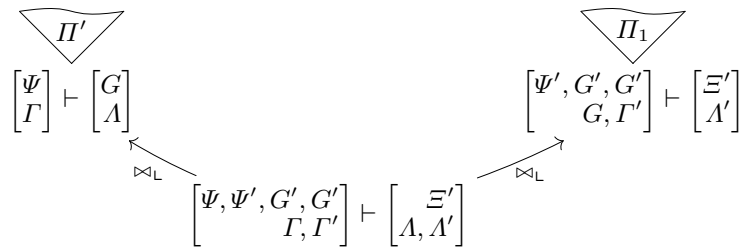

and obtain $\bar{\Pi}_1$. Then take $\bar{\Pi}$ as

$$
\dfrac{\begin{array}{c}\bar{\Pi}_1\\[2pt]\begin{bmatrix}\Psi,\Psi',G',G'\\\Gamma,\Gamma'\end{bmatrix}\vdash\begin{bmatrix}\Xi'\\\Lambda,\Lambda'\end{bmatrix}\end{array}}{\begin{bmatrix}\Psi,\Psi',G'\\\Gamma,\Gamma'\end{bmatrix}\vdash\begin{bmatrix}\Xi'\\\Lambda,\Lambda'\end{bmatrix}}\;{>}{\uparrow}\quad.
$$

7)   $\bowtie_{\mathsf{L}}$, Any rule. Let $\Pi$ be

$$
\dfrac{
\dfrac{\dfrac{\Pi_1}{\begin{bmatrix}\Psi_1\\\Gamma_1\end{bmatrix}\vdash\begin{bmatrix}G'\\\Lambda_1\end{bmatrix}}\quad\dfrac{\Pi_2}{\begin{bmatrix}\Psi_2\\G',\Gamma_2\end{bmatrix}\vdash\begin{bmatrix}G\\\Lambda_2\end{bmatrix}}}{\begin{bmatrix}\Psi_1,\Psi_2\\\Gamma_1,\Gamma_2\end{bmatrix}\vdash\begin{bmatrix}G\\\Lambda_1,\Lambda_2\end{bmatrix}}\;\bowtie_{\mathsf{L}}\qquad\dfrac{\Pi''}{\begin{bmatrix}\Psi'\\G,\Gamma'\end{bmatrix}\vdash\begin{bmatrix}\Xi'\\\Lambda'\end{bmatrix}}
}{\begin{bmatrix}\Psi_1,\Psi_2,\Psi'\\\Gamma_1,\Gamma_2,\Gamma'\end{bmatrix}\vdash\begin{bmatrix}\Xi'\\\Lambda_1,\Lambda_2,\Lambda'\end{bmatrix}}\;\bowtie_{\mathsf{L}}\uparrow\qquad.
$$

Apply the induction hypothesis on

$$
\dfrac{\dfrac{\Pi_2}{\begin{bmatrix}\Psi_2\\G',\Gamma_2\end{bmatrix}\vdash\begin{bmatrix}G\\\Lambda_2\end{bmatrix}}\qquad\dfrac{\Pi''}{\begin{bmatrix}\Psi'\\G,\Gamma'\end{bmatrix}\vdash\begin{bmatrix}\Xi'\\\Lambda'\end{bmatrix}}}{\begin{bmatrix}\Psi_2,\Psi'\\G',\Gamma_2,\Gamma'\end{bmatrix}\vdash\begin{bmatrix}\Xi'\\\Lambda_2,\Lambda'\end{bmatrix}}\;\bowtie_{\mathsf{L}}\uparrow
$$

and obtain $\bar{\Pi}_2$. Then take $\bar{\Pi}$ as

$$
\dfrac{\dfrac{\Pi_1}{\begin{bmatrix}\Psi_1\\\Gamma_1\end{bmatrix}\vdash\begin{bmatrix}G'\\\Lambda_1\end{bmatrix}}\qquad\dfrac{\bar{\Pi}_2}{\begin{bmatrix}\Psi_2,\Psi'\\G',\Gamma_2,\Gamma'\end{bmatrix}\vdash\begin{bmatrix}\Xi'\\\Lambda_2,\Lambda'\end{bmatrix}}}{\begin{bmatrix}\Psi_1,\Psi_2,\Psi'\\\Gamma_1,\Gamma_2,\Gamma'\end{bmatrix}\vdash\begin{bmatrix}\Xi'\\\Lambda_1,\Lambda_2,\Lambda'\end{bmatrix}}\;\bowtie_{\mathsf{L}}\uparrow\qquad.
$$

8)   $>$, any rule. Let $\Pi$ be

$$
\dfrac{
\dfrac{\dfrac{\dfrac{\Pi_1}{\begin{bmatrix}\Psi,G',G'\\\Gamma\end{bmatrix}\vdash\begin{bmatrix}G\\\Lambda\end{bmatrix}}}{\begin{bmatrix}\Psi,G'\\\Gamma\end{bmatrix}\vdash\begin{bmatrix}G\\\Lambda\end{bmatrix}}\;{>}{\uparrow}\qquad\dfrac{\Pi''}{\begin{bmatrix}\Psi'\\G,\Gamma'\end{bmatrix}\vdash\begin{bmatrix}\Xi'\\\Lambda'\end{bmatrix}}}{\begin{bmatrix}\Psi,\Psi',G'\\\Gamma,\Gamma'\end{bmatrix}\vdash\begin{bmatrix}\Xi'\\\Lambda,\Lambda'\end{bmatrix}}\;\bowtie_{\mathsf{L}}\uparrow
}{}\qquad.
$$

Apply the induction hypothesis on

$$
\frac{\overset{\Pi_1}{\nabla}}{\left[\begin{smallmatrix}\Psi, G', G'\\\Gamma\end{smallmatrix}\right] \vdash \left[\begin{smallmatrix}G\\\Lambda\end{smallmatrix}\right]} \qquad\qquad \frac{\overset{\Pi''}{\nabla}}{\left[\begin{smallmatrix}\Psi'\\G,\Gamma'\end{smallmatrix}\right] \vdash \left[\begin{smallmatrix}\Xi'\\\Lambda'\end{smallmatrix}\right]}
$$

$$
{}_{\bowtie_{\mathsf{L}}}\!\nearrow \qquad\qquad \nwarrow{}_{\bowtie_{\mathsf{L}}}
$$

$$
\left[\begin{smallmatrix}\Psi, \Psi', G', G'\\\Gamma, \Gamma'\end{smallmatrix}\right] \vdash \left[\begin{smallmatrix}\Xi'\\\Lambda, \Lambda'\end{smallmatrix}\right]
$$

and obtain $\bar{\Pi}_1$. Then take $\bar{\Pi}$ as

$$
\frac{\overset{\bar{\Pi}_1}{\nabla}}{\left[\begin{smallmatrix}\Psi, \Psi', G', G'\\\Gamma, \Gamma'\end{smallmatrix}\right] \vdash \left[\begin{smallmatrix}\Xi'\\\Lambda, \Lambda'\end{smallmatrix}\right]}
$$
$$
{}_{>}\!\uparrow
$$
$$
\left[\begin{smallmatrix}\Psi, \Psi', G'\\\Gamma, \Gamma'\end{smallmatrix}\right] \vdash \left[\begin{smallmatrix}\Xi'\\\Lambda, \Lambda'\end{smallmatrix}\right]
$$

.

□

**4.2.3 Lemma**   *For every proof in* G-Forum$^{>,\bowtie_{\mathsf{L}}}$ *there is a proof in* G-Forum$^{>}$ *whose conclusion is the same.*

**Proof**   Given a proof in G-Forum$^{>,\bowtie_{\mathsf{L}}}$, we say that a certain $\bowtie_{\mathsf{L}}$ instance is *maximal* if its cut-rank is higher than that of any other $\bowtie_{\mathsf{L}}$ instance above it.   Apply repeatedly Lemma 4.2.2 on maximal $\bowtie_{\mathsf{L}}$ instances until all $\bowtie_{\mathsf{L}}$ instances disappear: they do because the cut-rank in the subproofs whose root is a maximal $\bowtie_{\mathsf{L}}$ decreases each time.                     □

## 4.3   Elimination of the Contraction Rule

**4.3.1 Lemma**   *In* G-Forum$^{>}$, *let $\Pi$ be the proof*

$$
\frac{\overset{\Pi'}{\nabla}}{\left[\begin{smallmatrix}\Psi, G, G\\\Gamma\end{smallmatrix}\right] \vdash \left[\begin{smallmatrix}\Xi\\\Lambda\end{smallmatrix}\right]}
$$
$$
{}_{>}\!\uparrow
$$
$$
\left[\begin{smallmatrix}\Psi, G\\\Gamma\end{smallmatrix}\right] \vdash \left[\begin{smallmatrix}\Xi\\\Lambda\end{smallmatrix}\right]
$$

,

*where no instance of $>$ appears in $\Pi'$. A proof $\bar{\Pi}$ exists in* G-Forum *whose conclusion is the same as that of $\Pi$.*

**Proof**   By induction on the depth of $\Pi'$.

**Basis Case**

If $\Pi$ is

$$
\overset{\circ}{{}_{r}\!\uparrow}
$$
$$
\left[\begin{smallmatrix}\Psi, G, G\\\Gamma\end{smallmatrix}\right] \vdash \left[\begin{smallmatrix}\Xi\\\Lambda\end{smallmatrix}\right]
$$
$$
{}_{>}\!\uparrow
$$
$$
\left[\begin{smallmatrix}\Psi, G\\\Gamma\end{smallmatrix}\right] \vdash \left[\begin{smallmatrix}\Xi\\\Lambda\end{smallmatrix}\right]
$$

,

then take

$$\bar{\Pi} = \begin{array}{c} \circ \\ r \uparrow \\ \left[\begin{matrix} \Psi, G \\ \Gamma \end{matrix}\right] \vdash \left[\begin{matrix} \Xi \\ \Lambda \end{matrix}\right] \end{array} \quad ,$$

where $r$ is $\mathsf{G_L}$ or $\mathsf{G_R}$.

**Inductive Cases**

1) $\mathsf{G_L}$. Let $\Pi$ be

$$
\begin{array}{c}
\overline{\Pi_1''} \qquad \overline{\Pi_{k''}''} \qquad \overline{\Pi_1'} \qquad \overline{\Pi_{k'}'} \\
\left[\begin{matrix}\Psi,G,G\end{matrix}\right]\vdash\left[G_1\right] \ \cdots \ \left[\begin{matrix}\Psi,G,G\end{matrix}\right]\vdash\left[G_{k''}\right] \ \left[\begin{matrix}\Psi,G,G\\\Gamma_1\end{matrix}\right]\vdash\left[\begin{matrix}H_1\\\Lambda_1\end{matrix}\right] \ \cdots \ \left[\begin{matrix}\Psi,G,G\\\Gamma_{k'}\end{matrix}\right]\vdash\left[\begin{matrix}H_{k'}\\\Lambda_{k'}\end{matrix}\right] \\
\mathsf{G_L} \quad\quad \uparrow \mathsf{G_L} \quad\quad \mathsf{G_L} \\
\left[\begin{matrix}\Psi,G,G\\\Gamma\end{matrix}\right]\vdash\left[\Lambda\right] \\
> \uparrow \\
\left[\begin{matrix}\Psi,G\\\Gamma\end{matrix}\right]\vdash\left[\Lambda\right]
\end{array}
$$

where $k' + k'' > 0$. By the induction hypothesis, for $1 \leq j \leq k''$ and $1 \leq i \leq k'$, there are $>$-free proofs $\bar{\Pi}_j''$ and $\bar{\Pi}_i'$, corresponding respectively to

$$
\begin{array}{ccc}
\overline{\Pi_j''} & & \overline{\Pi_i'} \\
\left[\begin{matrix}\Psi,G,G\end{matrix}\right]\vdash\left[G_j\right] & \text{and} & \left[\begin{matrix}\Psi,G,G\\\Gamma_i\end{matrix}\right]\vdash\left[\begin{matrix}H_i\\\Lambda_i\end{matrix}\right] \\
> \uparrow & & > \uparrow \\
\left[\begin{matrix}\Psi,G\end{matrix}\right]\vdash\left[G_j\right] & & \left[\begin{matrix}\Psi,G\\\Gamma_i\end{matrix}\right]\vdash\left[\begin{matrix}H_i\\\Lambda_i\end{matrix}\right]
\end{array} \quad .
$$

Then take $\bar{\Pi}$ as

$$
\begin{array}{c}
\widehat{\bar{\Pi}_1''} \qquad \widehat{\bar{\Pi}_{k''}''} \qquad \widehat{\bar{\Pi}_1'} \qquad \widehat{\bar{\Pi}_{k'}'} \\
\left[\begin{matrix}\Psi,G\end{matrix}\right]\vdash\left[G_1\right] \ \cdots \ \left[\begin{matrix}\Psi,G\end{matrix}\right]\vdash\left[G_{k''}\right] \ \left[\begin{matrix}\Psi,G\\\Gamma_1\end{matrix}\right]\vdash\left[\begin{matrix}H_1\\\Lambda_1\end{matrix}\right] \ \cdots \ \left[\begin{matrix}\Psi,G\\\Gamma_{k'}\end{matrix}\right]\vdash\left[\begin{matrix}H_{k'}\\\Lambda_{k'}\end{matrix}\right] \\
\mathsf{G_L} \quad\quad \uparrow \mathsf{G_L} \quad\quad \mathsf{G_L} \\
\left[\begin{matrix}\Psi,G\\\Gamma\end{matrix}\right]\vdash\left[\Lambda\right]
\end{array} \quad .
$$

2) $\mathsf{G_R}$. Let $\Pi$ be

$$
\begin{array}{c}
\Pi_1 \qquad\qquad\qquad\qquad \Pi_h \\
\left[\begin{matrix}\Psi,G,G,\mathsf{cp}(\delta_1\rho)\\\Gamma,\mathsf{lp}(\delta_1\rho)\end{matrix}\right]\vdash\left[\mathsf{hd}(\delta_1\rho),\Lambda\right] \quad \cdots \quad \left[\begin{matrix}\Psi,G,G,\mathsf{cp}(\delta_h\rho)\\\Gamma,\mathsf{lp}(\delta_h\rho)\end{matrix}\right]\vdash\left[\mathsf{hd}(\delta_h\rho),\Lambda\right] \\
\mathsf{G_R} \qquad\qquad \mathsf{G_R} \\
\left[\begin{matrix}\Psi,G,G\\\Gamma\end{matrix}\right]\vdash\left[\begin{matrix}G'\\\Lambda\end{matrix}\right] \\
> \uparrow \\
\left[\begin{matrix}\Psi,G\\\Gamma\end{matrix}\right]\vdash\left[\begin{matrix}G'\\\Lambda\end{matrix}\right]
\end{array}
$$

where $h > 0$. By the induction hypothesis, for $1 \leq i \leq h$, there are $>$-free proofs $\bar{\Pi}_i$, corresponding to

$$
\overbrace{\phantom{xxxxx}}^{\Pi_i}
$$

$$
\left[\begin{matrix} \Psi, G, G, \mathsf{cp}(\delta_i\rho) \\ \Gamma, \mathsf{lp}(\delta_i\rho) \end{matrix}\right] \vdash \left[\mathsf{hd}(\delta_i\rho), \Lambda\right]
$$

$$
{}_{>}\Big\uparrow
$$

$$
\left[\begin{matrix} \Psi, G, \mathsf{cp}(\delta_i\rho) \\ \Gamma, \mathsf{lp}(\delta_i\rho) \end{matrix}\right] \vdash \left[\mathsf{hd}(\delta_i\rho), \Lambda\right]
$$

.

Then take $\bar{\Pi}$ as

$$
\overbrace{\phantom{xxx}}^{\bar{\Pi}_1} \qquad\qquad\qquad \overbrace{\phantom{xxx}}^{\bar{\Pi}_h}
$$

$$
\left[\begin{matrix} \Psi, G, \mathsf{cp}(\delta_1\rho) \\ \Gamma, \mathsf{lp}(\delta_1\rho) \end{matrix}\right] \vdash \left[\mathsf{hd}(\delta_1\rho), \Lambda\right] \quad \ldots \quad \left[\begin{matrix} \Psi, G, \mathsf{cp}(\delta_h\rho) \\ \Gamma, \mathsf{lp}(\delta_h\rho) \end{matrix}\right] \vdash \left[\mathsf{hd}(\delta_h\rho), \Lambda\right]
$$

$$
{}_{\mathsf{G_R}}\nwarrow \qquad \nearrow {}_{\mathsf{G_R}}
$$

$$
\left[\begin{matrix} \Psi, G \\ \Gamma \end{matrix}\right] \vdash \left[\begin{matrix} G' \\ \Lambda \end{matrix}\right]
$$

.

□

**4.3.2 Lemma**   *For every proof in* G-Forum$^>$ *there is a proof in* G-Forum *whose conclusion is the same.*

**Proof**   Apply repeatedly Lemma 4.3.1 until all $>$ instances disappear.                    □

## 4.4   The Cut Elimination Theorem

**4.4.1 Theorem**   *For every proof in* G-Forum$^{\bowtie_L,\bowtie_C}$ *there exists a proof in* G-Forum *with the same conclusion.*

**Proof**   G-Forum$^{>,\bowtie_L,\bowtie_C'}$ is more general than G-Forum$^{\bowtie_L,\bowtie_C}$, so we can proceed by the scheme:

$$
\text{G-Forum}^{>,\bowtie_L,\bowtie_C'} \to \text{G-Forum}^{>,\bowtie_L} \to \text{G-Forum}^> \to \text{G-Forum},
$$

by applying successively Lemma 4.1.2, 4.2.3 and 4.3.2.                    □

## 5   Conclusions

In this paper we showed G-Forum, a non-left-right symmetric sequent system for linear logic, and we proved cut elimination for it. Like in the case of Forum, the left-right asymmetry of sequents is motivated by the necessity of limiting proof search to uniform proofs. We limited the freedom of composing formulae in such a way that their structure matches that of sequents; in a certain sense, formulae become asymmetric, too. We consider this situation where two asymmetries match aestethically pleasant, and actually more symmetric than the same in Forum, where the structure of formulae is at odds with that of sequents. This new asymmetry is motivated by the desire of structuring proofs by easily definable, big building blocks, suitable to semantic understanding.

The result is a system for which it is natural to define cut rules and for which it is possible to prove cut elimination by a procedure that rewrites proofs inside the system, without resorting to Forum or plain linear logic. This guarantees that the new system has a good proof theoretical standing, which usually means that it is a good basis for further, fruitful research.

In a forthcoming paper we will show how to associate to G-Forum a labelled event structure semantics, i.e., a behavioural model of computation, along the lines initiated in [6]. In another paper we will apply G-Forum and its semantics to problems of partial order planning.

The methods in this paper are about studying the structure of proofs at a coarser abstraction level than the one provided by the sequent calculus. In another research project we are pursuing about the calculus of structures [3, 7, 2, 12] (see also http://alessio.guglielmi.name/res/cos), we study proofs at a *finer* level than provided by the sequent calculus. We do so for exploring properties of locality and modularity, which are important for concurrency, that are otherwise not available.

In the future, we plan to adapt the techniques in this paper to the calculus of structures (for example, of linear logic), in order to cover the full range of granularity: from the finer, suitable for distributed implementation, to the coarser, suitable for semantics.